\documentclass[aps,prd,amsfonts,groupedaddress,showpacs,showkeys]{revtex4}
\usepackage{latexsym,epsfig,amsmath,amssymb}

\usepackage{epsfig}
\usepackage{amssymb}
\newcommand{\seq}{\begin{subequations}}
\newcommand{\sen}{\end{subequations}} 
\newcommand{\eq}{\begin{eqnarray}}
\newcommand{\en}{\end{eqnarray}}
\newcommand{\ra}{\rangle}
\newcommand{\la}{\langle}

\def\bwt{\begin{widetext}}
\def\ewt{\end{widetext}}
\def\be{\begin{equation}}
\def\ee{\end{equation}}
\def\bea{\begin{eqnarray}}
\def\eea{\end{eqnarray}}
\def\bean{\begin{eqnarray*}}
\def\eean{\end{eqnarray*}}
\def\bary{\begin{array}}
\def\eary{\end{array}}

\def\bit{\begin{itemize}}
\def\eit{\end{itemize}}

\begin{document}

\title{Molecular structure of the 
$B_{s0}^{\ast}(5725)$ and $B_{s1}(5778)$ bottom-strange mesons} 
\author{
Amand Faessler,
Thomas Gutsche,
Valery E. Lyubovitskij
\footnote{On leave of absence from the
Department of Physics, Tomsk State University,
634050 Tomsk, Russia},
Yong-Liang Ma
\vspace*{1.2\baselineskip}}
\affiliation{Institut f\"ur Theoretische Physik,
Universit\"at T\"ubingen, \\ 
Auf der Morgenstelle 14, D--72076 T\"ubingen, Germany
\vspace*{0.3\baselineskip}\\}

\date{\today}

\begin{abstract} 

We discuss a possible interpretation of the 
scalar $B_{s0}^{\ast}(5725)$ and axial $B_{s1}(5778)$ 
bottom--strange mesons as hadronic molecules --- bound states 
of $B K$ and $B^\ast K$ mesons, respectively. 
Using a phenomenological Lagrangian approach we analyze 
the strong $B_{s0}^{\ast} \to B_s \pi^0$, 
$B_{s1} \to B_s^{\ast} \pi^0$ and the radiative 
$B_{s0}^{\ast} \to B_s^{\ast} \gamma$, $B_{s1} \to B_s \gamma$, 
$B_{s1} \to B_s^{\ast} \gamma$, $B_{s1} \to B_{s0}^{\ast} \gamma$ 
decays. We give predictions for the decay properties: 
effective couplings and decay widths. 
\end{abstract}



\maketitle

\newpage 

\section{Introduction}

Newly observed hadrons, in particular in the heavy flavor 
sector, have raised strong interest in treating and 
interpreting these states as hadronic bound states or, 
as commonly named, hadronic molecules (for overview see 
e.g. Ref.~\cite{Rosner:2006vc}). One of the main reasons 
to treat these observed states as molecules is that their 
masses are close to the thresholds of corresponding hadronic 
pairs. Canonical examples are the scalar $D_{s0}^\ast(2317)$
and axial $D_{s1}(2460)$ mesons. As stressed 
in~\cite{Barnes:2003dj} the scalar $D_{s0}^\ast(2317)$
and axial $D_{s1}(2460)$ mesons could be candidates for 
a scalar $DK$ and a axial $D^\ast K$ molecule because of 
a relatively small binding energy of $\sim 50$ MeV.

In a series of papers~\cite{Faessler:2007gv}-\cite{Dong:2007px} 
we developed the quantum field approach based on phenomenological 
Lagrangians for the treatment of hadrons as bound states of lighter 
hadrons --- hadronic molecules. In particular, we considered the 
strong, radiative and leptonic decays of $D_{s0}^\ast(2317)$ and 
$D_{s1}(2460)$ mesons~\cite{Faessler:2007gv}-\cite{Faessler:2007us}. 
A new feature related to the $DK$ and $D^\ast K$ molecular structure 
of the $D_{s0}^\ast$ and $D_{s1}$ mesons was that the 
presence of $u(d)$ quarks in the $D^{(\ast)}$ and $K$ mesons gives 
rise to direct strong isospin-violating transitions 
$D_{s0}^{\ast} \to D_s \pi^0$ and $D_{s1} \to D_s^\ast \pi^0$ in
addition to the decay mechanism induced by $\eta-\pi^0$ mixing, 
as considered previously. We showed that the direct transition
compares with and even dominates over the $\eta-\pi^0$ mixing 
transition in the isospin-violating decays of $D_{s0}^\ast$ and 
$D_{s1}$ mesons. The composite (molecular) structure of the 
$D_{s0}^\ast$ and $D_{s1}$ mesons was defined by the compositeness
condition $Z=0$~\cite{Weinberg:1962hj,Efimov:1993ei,Anikin:1995cf}
(see also Refs.~\cite{Baru:2003qq} 
and \cite{Faessler:2007gv}-\cite{Dong:2007px}). 
This condition implies that the renormalization constant of the hadron 
wave function is set equal to zero or that the hadron exists as 
a bound state of its constituents. The compositeness condition was 
originally applied to the study of the deuteron as a bound state of 
proton and neutron~\cite{Weinberg:1962hj}. Then it was extensively used
in low-energy hadron phenomenology as the master equation for the
treatment of mesons and baryons as bound states of light and heavy
constituent quarks (see e.g. Refs.~\cite{Efimov:1993ei,Anikin:1995cf}). 
We found that our theoretical framework gives numerical results 
which are consistent with both the experimental data and previous 
theoretical approaches. 

In this paper we extend our 
formalism~\cite{Faessler:2007gv}-\cite{Dong:2007px} to the 
scalar $B_{s0}^{\ast}(5725)$ and axial $B_{s1}(5778)$ 
bottom-strange mesons which have been considered 
before theoretically~\cite{Godfrey:1986wj}-\cite{Wang:2008bt} 
as the bottom partners of the charm mesons $D_{s0}^\ast$ and 
$D_{s1}$. In Ref.~\cite{Bardeen:2003kt} masses, strong and 
radiative decays of $B_{s0}^{\ast}$ and $B_{s1}$ states have been 
analyzed using heavy hadron chiral perturbation theory. In particular, 
the mass spectrum, strong and radiative decays widths have been 
calculated using quark models, QCD sum rules, effective field 
approaches based on chiral and heavy quark symmetry including 
coupled-channel unitarity. 

We assume that the $B_{s0}^{\ast}(5725)$ and $B_{s1}(5778)$ are bound 
states of $B K$ and $B^\ast K$ mesons, respectively. 
We adopt that the isospin, spin and parity quantum numbers of
the $B_{s0}^\ast$ ($\bar B_{s0}^\ast$) and $B_{s1}$ ($\bar B_{s1}$)  
are $I(J^P) = 0(0^+)$ and $I(J^P) = 0(1^+)$, while for their  masses 
we take the values $m_{B_{s0}^{\ast}} = 5725$ MeV and 
$m_{B_{s1}} = 5778$ MeV (the central values predicted in 
Refs.~\cite{Guo:2006fu,Guo:2006rp}). Note, that different approaches 
result in the following ranges  for the $B_{s0}^\ast$ and $B_{s1}$ 
masses: $m_{B_{s0}^{\ast}}$~= 5627 -- 5841 MeV and 
$m_{B_{s1}}$ = 5660 -- 5859 MeV. 
Using a phenomenological Lagrangian approach we analyze 
the strong $B_{s0}^{\ast} \to B_s \pi^0$, 
$B_{s1} \to B_s^{\ast} \pi^0$ and the radiative 
$B_{s0}^{\ast} \to B_s^{\ast} \gamma$, $B_{s1} \to B_s \gamma$,
$B_{s1} \to B_s^{\ast} \gamma$, $B_{s1} \to B_{s0}^{\ast} \gamma$ 
 decays. We give predictions for the decay properties: 
effective couplings and decay widths.  

In the present manuscript we proceed as follows. First, in 
Sec.~\ref{Frame}, we outline our framework. We discuss the effective
mesonic Lagrangian for the treatment of the $B_{s0}^\ast$ and $B_{s1}$ 
mesons as $B K$ and $B^\ast K$ bound states, respectively. In Section~III 
we consider the matrix elements describing the strong and radiative decays 
of the $B_{s0}^\ast$ and $B_{s1}$ mesons. We discuss our numerical results 
and perform a comparison with other theoretical approaches. 
In Section~IV we present a short summary of our results.

\section{Theoretical Framework}
\label{Frame}

\subsection{Molecular structure of the $B_{s0}^{\ast}(5725)$ 
and $B_{s1}(5778)$ mesons} 

In this section we discuss the formalism for the study of the
$B_{s0}^{\ast}(5725)$ and $B_{s1}(5778)$ mesons as hadronic 
molecules, represented by a $B K$ and $B^\ast K$ bound 
state, respectively. We adopt that their quantum numbers 
(isospin, spin and parity) are $I(J^P) = 0(0^+)$ 
for $B_{s0}^\ast$ ($\bar B_{s0}^\ast$) and $I(J^P) = 0(1^+)$ 
for $B_{s1}$ ($\bar B_{s1}$). For their masses we take the values 
$m_{B_{s0}^{\ast}} = 5725$ MeV and 
$m_{B_{s1}} = 5778$ MeV (the central values predicted in 
Refs.~\cite{Guo:2006fu,Guo:2006rp}). Our framework is based on an 
effective interaction Lagrangian describing the couplings of the 
$B_{s0}^\ast$ and $B_{s1}$ mesons to their constituents:
\seq\label{Bs0_Bs1_L} 
\eq 
{\cal L}_{B_{s0}^{\ast}} &=& g_{_{B_{s0}^\ast}} \, 
\bar{B}_{s0}^{\ast}(x)  \, \int\! dy \, \Phi_{B_{s0}^{\ast}}(y^2) \, 
B^T(x+\omega_{_{KB}}y) \bar K(x -\omega_{_{BK}} y) \, + \, {\rm H.c.} \,, \\
{\cal L}_{B_{s1}} &=& g_{_{B_{s1}}} \, \bar{B}_{s1}^{\mu}(x) 
\, \int\! dy \, \Phi_{B_{s1}}(y^2) \, 
B_\mu^{\ast \, T}(x+\omega_{_{KB^\ast}}y) \, \bar K(x - \omega_{_{B^\ast K}}y) 
\, + \, {\rm H.c.} \,, 
\end{eqnarray} 
\sen  
where the doublets of $B$, $B^\ast_\mu$ and $\bar K$ mesons are defined as 
\eq 
B = \left(
  \begin{array}{c}
    B^{+} \\
    B^{0} \\
  \end{array}
\right) \,, \,\,\,\,\, 
B_\mu^\ast = \left(
  \begin{array}{c}
    B^{\ast +} \\
    B^{\ast 0} \\
  \end{array}
\right)_\mu \,, \,\,\,\,\, 
\bar K = \left(
  \begin{array}{c}
    K^- \\
    \bar{K}^0 \\
  \end{array}
\right) \,. \label{Doublet}
\en 
The summation over isospin indices is understood and 
the symbol $T$ refers to the transpose of the doublets 
$B$ and $B^\ast$. The molecular structure of the $B_{s0}^\ast$ 
and $B_{s1}$ states is (we do not consider isospin mixing): 
$|B_{s0}^{\ast}\ra = |B^+K^-\ra + |B^0 \bar{K}^0\ra\,,$   
$|B_{s1}\ra = |B^{\ast +}K^-\ra + |B^{\ast 0}\bar{K}^0\ra$. 
In Eq.~(\ref{Bs0_Bs1_L}) we introduced the kinematical parameters 
$w_{ij} = m_i/(m_i + m_j)$, where $m_{i,j} = m_B$, $m_{B^\ast}$ and $m_K$ 
are the masses of $B$, $B^\ast$ and $K$ mesons. 
The correlation functions $\Phi_M$ with $M = B_{s0}^\ast$ or $B_{s1}$
characterize the finite size of the $B_{s0}^\ast$ and $B_{s1}$
mesons as $BK$ and $B^\ast K$ bound states and depend on the relative
Jacobi coordinate $y$ with, in addition, $x$ being the center of mass
(CM) coordinate. Note, that the local limit corresponds to the substitution
of $\Phi_M$ by the Dirac delta-function: $\Phi_M(y^2) \to \delta^4 (y)$. 
A basic requirement for the choice of an explicit form of the correlation
function is that its Fourier transform vanishes sufficiently fast in the
ultraviolet region of Euclidean space to render the Feynman diagrams
ultraviolet finite. We adopt the Gaussian form,
$\tilde\Phi_M(p_E^2/\Lambda_M^2) \doteq \exp( - p_E^2/\Lambda_M^2)\,,$
for the Fourier transform of vertex function, where $p_{E}$ is the
Euclidean Jacobi momentum. Here $\Lambda_{B_{s0}^\ast}$
is a size parameter, which parametrizes the distribution of
$B$ and $K$ mesons inside the $B_{s0}^\ast$ molecule, while
$\Lambda_{B_{s1}}$ is the size parameter for the $B_{s1}$ molecule.
For simplicity we will use a universal scale parameter
$\Lambda_M = \Lambda_{B_{s0}^\ast} = \Lambda_{B_{s1}}$. 

\begin{figure}
\vspace*{2cm}
\begin{center}
\epsfig{file=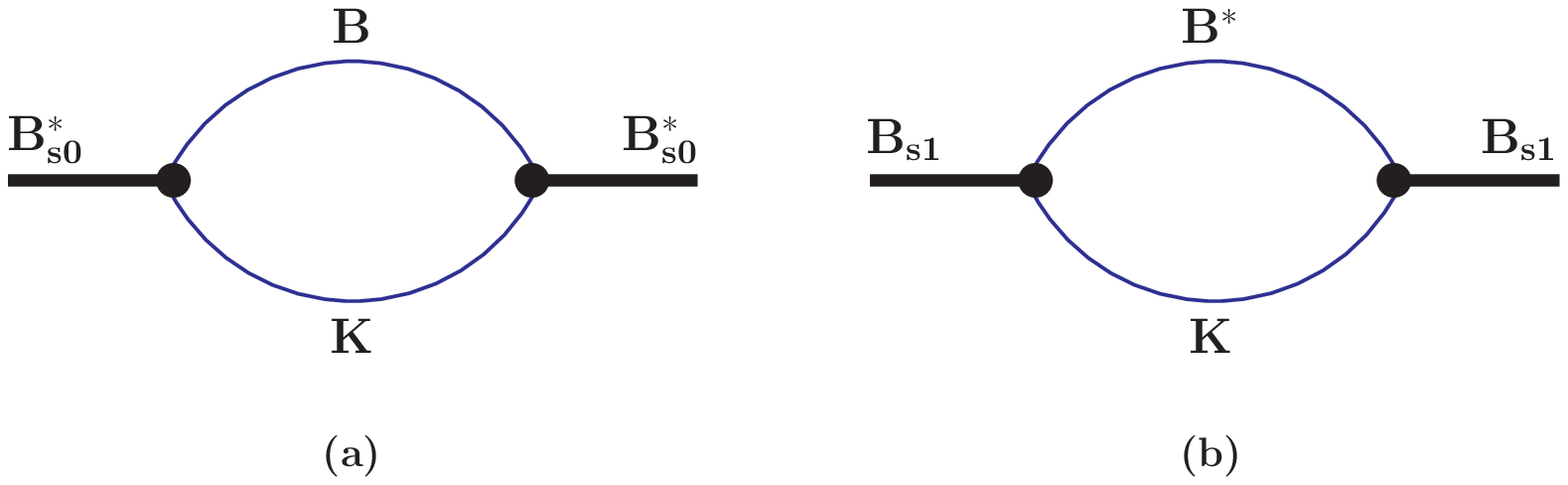, scale=.65}
\vspace*{1cm}
\caption{Mass operators of $B_{s0}^{\ast}$ and $B_{s1}$ mesons.}
\label{fig:Mass} 
\end{center}
\end{figure}

The coupling constants $g_{B_{s0}^\ast}$ and $g_{B_{s1}}$ are determined
by the compositeness condition~\cite{Weinberg:1962hj,Efimov:1993ei},
which implies that the renormalization constant of the hadron
wave function is set equal to zero:
\seq
\eq 
Z_{B_{s0}^\ast} &=& 1 -
\Sigma^\prime_{B_{s0}^\ast}(m_{B_{s0}^\ast}^2) = 0 \,, 
\label{non-renorm-1}\\
Z_{B_{s1}} &=& 1 - \Sigma^\prime_{B_{s1}}(m_{B_{s1}}^2) = 0 \,.
\label{non-renorm-2}
\en 
\sen 
Here, $\Sigma^\prime_{B_{s0}^\ast}(m_{B_{s0}^\ast}^2) =
g_{_{B_{s0}^\ast}}^2 \Pi^\prime_{B_{s0}^\ast}(m_{B_{s0}^\ast}^2)$
is the derivative of the $B_{s0}^\ast$ meson mass operator. 
In the case of the $B_{s1}$ meson we have
$\Sigma^\prime_{B_{s1}}(m_{B_{s1}}^2)
= g_{_{B_{s1}}}^2 \Pi^\prime_{B_{s1}}(m_{B_{s1}}^2)$, which is the
derivative of the transverse part of its mass operator
$\Sigma^{\mu\nu}_{B_{s1}}$, conventionally split into transverse
$\Sigma_{B_{s1}}$ and longitudinal $\Sigma^L_{B_{s1}}$  parts as:
\eq
\Sigma^{\mu\nu}_{B_{s1}}(p) =
g^{\mu\nu}_\perp \Sigma_{B_{s1}}(p^2) + \frac{p^\mu p^\nu}{p^2}
\Sigma^L_{B_{s1}}(p^2) \,,
\en
where
\eq
g^{\mu\nu}_\perp = g^{\mu\nu} - \frac{p^\mu p^\nu}{p^2}\,,
\hspace*{.5cm} g^{\mu\nu}_\perp p_\mu = 0\,.
\en
The mass operators of the $B_{s0}^\ast$ and $B_{s1}$ mesons
are described by the diagrams of Fig.~\ref{fig:Mass}(a)  
and~\ref{fig:Mass}(b), respectively.

Following Eqs.~(\ref{non-renorm-1}) and (\ref{non-renorm-2})
the coupling constants $g_{_{B_{s0}^\ast}}$ and $g_{_{B_{s1}}}$
can be expressed in the form:
\seq 
\eq
\frac{1}{g_{_{B_{s0}^\ast}}^2} &=& \frac{2}{(4 \pi \Lambda_M)^2} \,
\int\limits_0^1 dx \int\limits_0^\infty
\frac{d\alpha \, \alpha \, P_0(\alpha, x)}{(1 + \alpha)^3}
\,\, \biggl[ - \frac{d}{dz_0} \tilde \Phi^2_M(z_0) \biggr] \,,
\label{coupling_Bs0}\\
\frac{1}{g_{_{B_{s1}}}^2} &=& \frac{2}{(4 \pi \Lambda_M)^2} \,
\int\limits_0^1 dx \int\limits_0^\infty
\frac{d\alpha \, \alpha \, P_1(\alpha, x)}{(1 + \alpha)^3}
\,\, \biggl[ \frac{1}{2 \mu_{B^\ast}^2 (1 + \alpha)}
- \frac{d}{dz_1} \biggr] \tilde \Phi^2_M(z_1)
\label{coupling_Bs1}
\en 
\sen 
where
\eq
P_0(\alpha, x)
&=& \alpha^2 x(1-x) + w_{_{BK}}^2 \alpha x + w_{_{KB}}^2 \alpha (1-x)
\,, \nonumber\\
P_1(\alpha, x)
&=& \alpha^2 x(1-x) + w_{_{B^\ast \! K}}^2 \alpha x
+ w_{_{KB^\ast}}^2 \alpha (1-x) \,, \\
z_0 &=& \mu_B^2 \alpha x +  \mu_K^2 \alpha (1-x)
     - \frac{P_0(\alpha, x)}{1 + \alpha}  \, 
\mu_{B_{s0}^\ast}^2 \,, \nonumber\\
z_1 &=& \mu_{B^\ast}^2 \alpha x + \mu_K^2 \alpha (1-x)
   - \frac{P_1(\alpha, x)}{1 + \alpha}  \, \mu_{B_{s1}}^2 \,, \nonumber\\
\mu_M &=& \frac{m_M}{\Lambda_M} \,. \nonumber
\en
The above expressions are valid for any functional form of the correlation
function ${\widetilde{\Phi}}_M$. 

Note that the compositeness condition of the
type~(\ref{non-renorm-1}), (\ref{non-renorm-2}) was originally applied
to the study of the deuteron as a bound state of proton and
neutron~\cite{Weinberg:1962hj}. Then this condition was extensively used
in low-energy hadron phenomenology as the master equation for the
treatment of mesons and baryons as bound states of light and heavy
constituent quarks~\cite{Efimov:1993ei,Anikin:1995cf}. 
In Refs.~\cite{Baru:2003qq} and~\cite{Faessler:2007gv}-\cite{Dong:2007px} 
this condition was used in the application to hadronic molecule 
configurations of light and heavy mesons. 

\subsection{Effective Lagrangian for strong and radiative 
decays of $B_{s0}^\ast$ and $B_{s1}$ mesons} 

In Ref.~\cite{Faessler:2007gv,Faessler:2007us}, in the analysis of 
the strong isospin-violating decays $D_{s0}^\ast \to D_s \pi^0$  
and $D_{s1} \to D_s^\ast \pi^0$ in the molecular approach, we showed 
the existence of two possible  dynamical mechanisms. This included the 
so-called ``direct'' mechanism with $\pi^0$-meson emission from the 
$D \to D^\ast$ and $K \to K^\ast$ transitions and the ``indirect'' 
mechanism where a $\pi^0$ meson is produced via $\eta-\pi^0$ mixing.  
The mixing is due to the mass term of pseudoscalar mesons in the 
leading-order $O(p^2)$ Lagrangian of chiral perturbation 
theory (ChPT)~\cite{Gasser:1984gg,Cho:1994zu}.
Note, that the second mechanism based on $\eta-\pi^0$ mixing
was mainly considered before in the literature. Originally 
it was initiated by the analysis based on the use of chiral
Lagrangians~\cite{Cho:1994zu,Bardeen:2003kt,Colangelo:2003vg,Mehen:2004uj} 
where the leading-order, tree-level 
$D_{s0}^\ast D_s \pi^0$ ($D_{s1} D_s^\ast \pi^0$) coupling
can be generated only by virtual $\eta$-meson emission. 
In Ref.~\cite{Faessler:2007us} we showed that the ``direct'' and 
``mixing'' mechanisms can be combined together in the form of an 
effective coupling of $\pi^0$ to the mesonic pairs $DD^\ast$ or 
$KK^\ast$ with modified flavor structure. This modification occurs 
after the diagonalization of the mesonic mass term involving $\pi^0$ 
and $\eta$ meson fields~\cite{Gasser:1984gg} 
(see details in~\cite{Faessler:2007us}). 
In particular, instead of the $\tau_3 \, \pi^0$ coupling to
$DD^\ast$ or $KK^\ast$ we have
$\pi^0 \, ( \tau_3 \, \cos\varepsilon \, + \, \kappa \, I \,
\sin\varepsilon)$, where $\kappa = 1/\sqrt{3}$ or $\sqrt{3}$ is
the corresponding flavor-algebra factor for the $DD^\ast$ or 
$KK^\ast$ coupling, respectively. The $\pi^0-\eta$ mixing angle 
$\varepsilon$ is fixed as~\cite{Gasser:1984gg}:  
\eq  
\tan 2\varepsilon = \frac{\sqrt{3}}{2} \frac{m_d - m_u}{m_s - \hat{m}} 
\simeq 0.02  \,, \hspace*{1cm} \hat{m} = \frac{1}{2} (m_u + m_d)\,, 
\en
where $m_u, m_d, m_s$ are the current quark masses. 

Below, in Eq.(\ref{L_strong}), 
we display the explicit form of the corresponding 
interaction Lagrangian. The lowest-order diagrams which contribute 
to the matrix elements of the strong isospin-violating decays 
$B_{s0}^\ast \to B_s \pi^0$ and $B_{s1} \to B_s^\ast \pi^0$ are shown 
in Figs.2 and 3. 
Note, that in the isospin limit ($m_u = m_d$), the $\eta-\pi^0$
mixing angle vanishes and the masses of the virtual $B^{(\ast)}$ and
$K^{(\ast)}$ mesons in the loops are degenerate. 
As a result the pairs of diagrams related to
Figs.2(a) and 2(b), Figs.2(c) and 2(d), 
Figs.3(a) and 3(b), Figs.3(c) and 3(d) compensate each other. Therefore,
in the calculation of the diagrams of Figs.2 and 3 we go beyond the isospin 
limit and use the physical meson masses.

The diagrams contributing to the radiative decays 
$B_{s0}^\ast \to B_s^{\ast} \gamma$, $B_{s1} \to B_s \gamma$, 
$B_{s1} \to B_s^{\ast} \gamma$ and $B_{s1} \to B_{s0}^{\ast} \gamma$  
are shown in Figs.4, 5, 6 and 7.  
The diagrams of Figs.4(a), 4(b), 5(a) and 5(b) are generated by the 
direct coupling of the charged $K^-$ and $B^+ (B^{\ast +})$ mesons to 
the electromagnetic field after gauging the free Lagrangians related 
to these mesons. The diagrams of Figs.6(a)-6(d), 7(a) and 7(b) are 
generated by the coupling of a corresponding 
pair of vector and pseudoscalar mesons to the photon.  
The diagrams of Figs.4(c), 5(c), and 6(e) are generated 
after gauging the nonlocal strong Lagrangians~(\ref{Bs0_Bs1_L}) describing 
the coupling of the $B_{s0}^\ast$ and $B_{s1}$ mesons to their 
constituents. The diagrams of Figs.4(d) 
and 5(d) arise after gauging the strong $B_s^{\ast} B^+ K^-$ and 
$B_s B^{\ast +} K^-$ interaction Lagrangian containing
derivatives acting on the pseudoscalar fields. 
Details of how to generate the effective couplings of the involved
mesons to the electromagnetic field will be discussed later.
 
After the preliminary discussion of the relevant diagrams, we are now 
in the position to write down the full effective Lagrangian for the 
study of the strong and radiative decays of the $B_{s0}^\ast$ and 
$B_{s1}$ mesons formulated in terms of mesonic degrees of freedom and 
of photons. We follow the procedure discussed in detail in 
Refs.~\cite{Faessler:2007gv}-\cite{Faessler:2007us}, where we 
considered the $D_{s0}^\ast$ and $D_{s1}$ meson decay properties. 
First, we write the Lagrangian ${\cal L}$, which includes the free 
mesonic parts ${\cal L}_{\rm free}$ and the strong interaction parts 
${\cal L}_{\rm str}$: 
\eq\label{L_full} 
{\cal L}(x) \, = \, 
{\cal L}_{\rm free}(x) \, + \, {\cal L}_{\rm str}(x) \,,
\en 
with
\eq\label{L_free} 
{\cal L}_{\rm free}(x) &=& 
- \bar B_{s0}^{\ast\, +} (x) ( \Box + m_{B_{s0}^\ast}^2 )  
B_{s0}^{\ast}(x) 
+ \bar B_{s1, \mu} (x) ( g^{\mu\nu} [\Box + m_{B_{s1}}^2] 
- \partial^\mu\partial^\nu ) B_{s1, \nu}(x) \nonumber\\[2mm]
&-&\frac{1}{2} \vec\pi(x) ( \Box + m_\pi^2 ) \vec{\pi}(x) 
+ \frac{\delta_\pi}{2} \, [\pi^0(x)]^2 - 
\sum\limits_{P = K, B, B_s} P^\dag(x) ( \Box + m_P^2 ) P(x) 
+ \sum\limits_{P = K, B} \delta_P \, \bar P^0(x) P^0(x) \nonumber\\[2mm]
&+&\sum\limits_{V = K^\ast, B^\ast, B_s^\ast}  
V^{\dagger}_\mu (x) ( g^{\mu\nu} [\Box + m_{V}^2] 
- \partial^\mu\partial^\nu ) V_\nu(x) 
- \sum\limits_{V = K^\ast, B^\ast} \delta_{V} \, \bar  V^0_\mu (x) 
V^{0 \, \mu} (x) \,, 
\en 
\eq\label{L_strong}   
{\cal L}_{\rm str}(x) & = & \frac{g_{_{B^\ast B \pi}}}{2\sqrt{2}} 
\, B^{\ast \, \dagger}_\mu(x) \, \hat{\pi}_B(x) \, 
i\!\stackrel{\leftrightarrow}{\partial}^{\,\mu} \!\! B(x) 
+ \frac{g_{_{K^\ast K \pi}}}{\sqrt{2}} \, K^{\ast \, \dagger}_\mu(x) \, 
\hat{\pi}_K(x) \, i\!\stackrel{\leftrightarrow}{\partial}^{\,\mu}\!\! K(x) 
\nonumber\\[1mm]  
&+& g_{_{B_s B^\ast K}} \, B^{\ast \, \dagger}_\mu(x) \, K(x)
\, i\!\stackrel{\leftrightarrow}{\partial}^{\,\mu} \!\! B_s^0(x) 
+ g_{_{B_s B K^\ast}} \, K^{\ast \, \dagger}_\mu(x) \, B(x)
\, i\!\stackrel{\leftrightarrow}{\partial}^{\,\mu} \!\! \bar B_s^0(x)
+ g_{_{B^\ast_s B K}} 
B^{\ast \, 0}_{s, \mu}(x) \, B^\dagger(x)
\, i\!\stackrel{\leftrightarrow}{\partial}^{\,\mu} \!\! K(x) 
\nonumber\\[1mm] 
&-& i g_{_{B_s^\ast B^\ast K^\ast}} \, 
\biggl[ B^{\ast \, 0 \, \mu\nu }_s(x) B^{\ast \, \dagger}_\mu(x) 
K^\ast_\nu(x) + B^{\ast \, \dagger}_{\mu\nu}(x) 
K^{\ast \, \mu}(x) B^{\ast \, 0 \, \nu}_{s}(x) + K^{\ast \, \mu\nu}(x) 
B^{\ast \, 0}_{s, \mu}(x) B^{\ast \, \dagger}_\nu(x) \biggr] 
\nonumber\\[1mm] 
&+& \frac{g_{B_s^\ast B^\ast K}}{4} \epsilon^{\mu\nu\alpha\beta} 
B^{\ast}_{s \, \mu\nu}(x) B^{\ast \, \dagger}_{\alpha\beta}(x) K(x) 
+ {\cal L}_{B_{s0}^{\ast}}(x) +  {\cal L}_{B_{s1}}(x) 
\, + \, {\rm H.c.} \,,  
\en 
where summation over isospin indices is understood, 
$\Box = \partial^\mu \partial_\mu$ and 
$A \stackrel{\leftrightarrow}{\partial} B 
\equiv A \partial B - B \partial A$. Here, $\vec\pi = (\pi_1, \pi_2, \pi_3)$ 
is the triplet of pions, $\hat{\pi}_B = \pi_1 \tau_1 + \pi_2 \tau_2 
+ \pi_3 (\tau_3 \cos\varepsilon + I \sin\varepsilon/\sqrt{3})$, 
$\hat{\pi}_K = \pi_1 \tau_1 + \pi_2 \tau_2 
+ \pi_3 (\tau_3 \cos\varepsilon + I \sin\varepsilon \sqrt{3})$, 
 $B^{(\ast)}$ and $K^{(\ast)}$ are the doublets of pseudoscalar (vector) 
mesons, $B_s^\pm$ and $B_s^{\ast \, \pm}$ 
are the pseudoscalar and vector bottom-strange mesons, respectively, 
$V^{\ast \, \mu\nu} = \partial^\mu V^{\ast \, \nu} -   
\partial^\nu V^{\ast \, \mu}$ is the stress tensor of the 
vector meson field.   

In our convention the isospin-symmetric meson masses of the 
isomultiplets $m_\pi, m_P, m_V$ are identified with the masses 
of the charged partners. The quantities $\delta_{M}$ are the 
isospin-breaking parameters which are fixed by the difference of masses 
squared of the charged and neutral members of the isomultiplets as:
$\delta_M  = m_{M^\pm}^2 - m_{M^0}^2$ and
$m_{M^0} \equiv m_{\bar M^0}\,.$  
The set of mesonic masses is taken from data~\cite{Yao:2006px}. 
From Eq.~(\ref{L_strong}) it is evident that the couplings of $\pi^0$ 
to the $B^\ast B$ and $K^\ast K$ mesonic pairs contain two terms ---   
the ``dominant'' coupling (proportional to $\cos\varepsilon$) and 
the ``suppressed'' coupling (proportional to $\sin\varepsilon$). 
This means that the first coupling survives in the isospin limit, 
while the second one vanishes. 
  
The free meson propagators are given by the standard expressions 
\eq
i \, D_M(x-y) = \la 0 | T \, M(x) \, M^\dagger(y)  | 0 \ra 
\ = \
\int\frac{d^4k}{(2\pi)^4i} \, e^{-ik(x-y)} \ \tilde D_M(k) 
\en 
for the scalar (pseudoscalar) fields, where 
$\tilde D_M(k) = (m_M^2 - k^2 - i\epsilon)^{-1}$
and 
\eq
i \, D_{M^\ast}^{\mu\nu}(x-y) = \la 0 | T \, M^{\ast \, \mu}(x) \,
M^{\ast \, \nu \, \dagger}(y) | 0 \ra \ = \
\int\frac{d^4k}{(2\pi)^4i} \, e^{-ik(x-y)} \
\tilde D_{M^\ast}^{\mu\nu}(k)
\en 
for the vector (axial) fields, where
$\tilde D_{M^\ast}^{\mu\nu}(k) = ( - g^{\mu\nu} + k^\mu k^\nu/m_{M^\ast}^2) \, 
(m_{M^\ast}^2 - k^2 - i\epsilon)^{-1}\,.$ 
The choice for the strong meson couplings 
of the Lagrangian (\ref{L_strong}) will be discussed in Sec.III.    

The electromagnetic field is included in the Lagrangian (\ref{L_full}) 
using minimal substitution i.e. each derivative acting on a charged meson 
field is replaced by the covariant one:  
$\partial^\mu M^{(\ast) \pm} \, \to \, 
(\partial^\mu \mp ie A^\mu) \, M^{(\ast) \pm} \,.$
Note, that the strong interaction Lagrangians 
${\cal L}_{B_{s0}^{\ast}}$ and ${\cal L}_{B_{s1}}$ should also be 
modified in order to restore electromagnetic gauge invariance. 
It proceeds in a way as suggested in Ref.~\cite{Mandelstam:1962mi} 
and is extensively used in 
Refs.~\cite{Anikin:1995cf,Faessler:2007gv,Faessler:2007us}. 
In particular, each charged constituent meson field $H^{\pm}$ 
(i.e. $B^{(\ast \pm)}$ and $K^{(\pm)}$) in 
${\cal L}_{B_{s0}^{\ast}}$ and ${\cal L}_{B_{s1}}$ 
is multiplied by the gauge field exponential (see further details 
in~\cite{Anikin:1995cf,Faessler:2007gv,Faessler:2007us}): 
\eq\label{Subs_nonmin} 
H^{\pm}(y) \to e^{\mp i e I(y,x,P)} H^{\pm}(y) 
\en 
where
\eq\label{path}
I(x,y,P) = \int\limits_y^x dz_\mu A^\mu(z).
\en
For the derivative of $I(x,y,P)$ we use the
path-independent prescription suggested in~\cite{Mandelstam:1962mi}
which in turn states that the derivative of $I(x,y,P)$ does
not depend on the path $P$ originally used in the definition. 
The nonminimal substitution (\ref{Subs_nonmin}) is therefore 
completely equivalent to the minimal prescription. Expanding the 
exponential $e^{\mp i e I(y,x,P)} H^{\pm}(y)$ in powers of 
the electromagnetic field and keeping linear terms like 
the four-particle coupling $B_{s0}^\ast B^+ K^- \gamma$ and 
$B_{s1} B^{\ast +} K^- \gamma$ we generate the diagrams of 
Figs.4(c) and 5(c), 6(c), 6(d), respectively.  

Finally, we specify the electromagnetic Lagrangian describing the coupling 
of vector and pseudoscalar mesons to the photon: 
\eq\label{L_em}  
{\cal L}_{V P \gamma}(x) &=& \frac{e}{4} \, F_{\mu\nu}(x) 
\, \epsilon^{\mu\nu\alpha\beta} \biggl( 
 g_{_{K^{\ast \, \pm} K^\pm \gamma}} 
 \, K^{\ast \, +}_{\alpha\beta}(x) \, K^-(x) 
+ g_{_{K^{\ast \, 0} K^0 \gamma}} 
 \, K^{\ast \, 0}_{\alpha\beta}(x) \, \bar K^0(x)\nonumber\\  
&+& g_{_{B^{\ast \, \pm } B^\pm \gamma}} 
 \, B^{\ast \, +}_{\alpha\beta}(x) \, B^-(x) 
+ g_{_{B^{\ast \, 0} B^0 \gamma}} 
 \, B^{\ast \, 0}_{\alpha\beta}(x) \, \bar B^0(x) \biggr)  
 \, + \, {\rm H.c.} 
\en 
Here the couplings $g_{_{V P \gamma}}$ can be extracted from the 
corresponding decay widths $V \to P + \gamma$. Presently we only 
have information about the decay width of $K^\ast$ mesons. 
Using the expressions for the $K^\ast \to K + \gamma$ decay widths 
\eq 
\Gamma(K^{\ast} \to K \gamma) = \frac{\alpha}{24} \, 
g_{_{K^\ast K^\gamma}}^2 \, m_{K^\ast}^3 \, 
\biggl( 1 - \frac{m_K^2}{m_{K^\ast}^2} \biggr)^3 \, 
\en 
and the data (central values) for 
$\Gamma(K^{\ast \, \pm} \to K^\pm \gamma) = 50.29$ keV and 
$\Gamma(K^{\ast \, 0} \to K^0 \gamma) = 116.19$ keV 
we deduce the coupling constants 
$g_{_{K^{\ast \, \pm} \to K^\pm \gamma}} = 0.836$ GeV$^{-1}$ 
and $g_{_{K^{\ast \, 0} \to K^0 \gamma}} = - 1.267$ GeV$^{-1}$. 
Note, that in the nonrelativistic SU(3) quark model 
the couplings $g_{_{K^{\ast \, \pm} \to K^\pm \gamma}}$ 
and $g_{_{K^{\ast \, 0} \to K^0 \gamma}}$ are proportional 
to the sum of the charges of the constituent quarks: 
$g_{_{K^{\ast \, \pm} \to K^\pm \gamma}} \sim (e_u + e_s) = 1/3$
and $g_{_{K^{\ast \, 0} \to K^0 \gamma}} \sim (e_d + e_s) = - 2/3$. 
This is the reason why the coupling of the neutral kaons is defined 
(by convention) with a negative sign. 
Also, the prediction of the nonrelativistic quark model for the ratio 
$g_{_{K^{\ast \, 0} \to K^0 \gamma}}/
g_{_{K^{\ast \, \pm} \to K^\pm \gamma}} = - 2$ is violated by relativistic 
corrections. For $D$ mesons the corresponding ratio is in precise 
agreement with data.   
In particular, taking the experimental values of 
$g_{_{D^{\ast \, \pm} \to D^\pm \gamma}} \simeq 0.5$ GeV$^{-1}$ 
and $g_{_{D^{\ast \, \pm} \to D^\pm \gamma}} \simeq 2.0$ GeV$^{-1}$ 
(see the discussion in Ref.~\cite{Dong:2008gb}) we 
get 
\eq 
\frac{g_{_{D^{\ast \, 0} \to D^0 \gamma}}}
{g_{_{D^{\ast \, \pm} \to D^\pm \gamma}}} = \frac{e_c + e_u}{e_c + e_d} 
= 4 \,. 
\en 
Therefore, one can expect that for bottom mesons the naive quark model 
should prediction should also be sufficient: 
\eq 
\frac{g_{_{B^{\ast \, 0} \to B^0 \gamma}}}
{g_{_{B^{\ast \, \pm} \to B^\pm \gamma}}} = \frac{e_b + e_d}{e_b + e_u} 
= - 2 \,. 
\en 
For our numerical estimates we will use typical values with
$g_{_{B^{\ast \, \pm} \to B^\pm \gamma}} =  0.5$ GeV$^{-1}$ 
and $g_{_{B^{\ast \, 0} \to B^0 \gamma}} =  - 1$ GeV$^{-1}$. 
These couplings correspond to the full width of $B^{\ast \, \pm}$ equal 
to 0.23 keV and and of $B^{\ast \, 0}$ equal to 0.91 keV (as usual 
we suppose that $B^\ast \to B \gamma$ is the dominant mode for the 
$B^\ast$ mesons). 

\section{Strong and radiative decays of the 
$B_{s0}^\ast$ and $B_{s1}$ mesons} 

\subsection{Matrix elements and decay widths} 

The matrix elements describing the strong 
$B_{s0}^\ast \to B_s \pi^0$, $B_{s1} \to B_s^\ast \pi^0$  
and radiative $B_{s0}^\ast \to B_s^\ast \gamma$, $B_{s1} \to B_s \gamma$ 
decays are defined as follows 
\seq\label{Minv_str} 
\eq 
M(B_{s0}^{\ast}(p) \to B_s(p^\prime) \pi^0(q)) 
&=& G_{B_{s0}^{\ast}B_s\pi} \,, \\
M(B_{s1}(p) \to B_s^\ast(p^\prime) \pi^0(q) ) &=&
\epsilon_\mu (p) \epsilon^{\ast}_\nu(p^\prime) \ 
( g^{\mu\nu} \ G_{B_{s1}B_s^\ast\pi}
\ - \ v^{\prime\mu} v^\nu \ F_{B_{s1}B_s^\ast\pi} ) \,, 
\en 
\sen 
and 
\seq\label{Minv_em}   
\eq
M(B_{s0}^{\ast}(p) \to B_s^{\ast}(p^\prime) \gamma(q)) 
&=&e \ \epsilon_\mu^\ast(q)\epsilon_\nu^\ast(p^\prime)\, 
( g_{\mu\nu} p^\prime q -
p^\prime_\mu q_\nu ) \ G_{B_{s0}^{\ast} B_s^{\ast} \gamma}
\,, \\ 
M( B_{s1}(p) \to B_s(p^\prime) \gamma(q) ) &=& 
e \ \epsilon_\mu(p)\epsilon^\ast_\nu(q) 
\ (g^{\mu\nu} \ pq \ - \ q^\mu p^\nu) \ G_{B_{s1}B_s\gamma}\,, \\
M( B_{s1}(p) \to B_s^\ast(p^\prime) \gamma(q) ) &=&
e \, \varepsilon^{m n\rho\sigma} \, \epsilon^\alpha(p)
\, \epsilon^{\ast \mu}(p^\prime) \, \epsilon_\rho^\ast(q) \, q_\sigma \, 
\biggl( G_{B_{s1}  B_s^{\ast} \gamma} \, g_{\mu n} g_{\alpha m} \, p q 
\nonumber\\
&+& \, F_{B_{s1}  B_s^{\ast} \gamma}  \, g_{\mu n} \,
p_m q_\alpha \, + \, H_{B_{s1}  B_s^{\ast} \gamma}  \, 
g_{\alpha m} \, p_n q_\mu\biggr) \,,\\ 
M( B_{s1}(p) \to B_{s0}^\ast(p^\prime) \gamma(q) ) &=& 
e \ \varepsilon^{\mu\nu\alpha\beta} \ \epsilon_\mu(p)\epsilon^\ast_\nu(q) 
\ p_\alpha \ q_\beta \ G_{B_{s1}B_{s0}\gamma}\,, 
\en 
\sen 
where $v = p/m_{B_{s1}}$ and $v^\prime = p^\prime/m_{B_s^\ast}$ 
are the four-velocities of the $B_{s1}$ and $B_s^\ast$ mesons, 
$G_{B_{s0}^{\ast}B_s\pi}$, $G(F)_{B_{s1}B_s^\ast\pi}$, 
$G_{B_{s0}^{\ast} B_s^{\ast} \gamma}$ and $G_{B_{s1}B_s\gamma}$, 
$G(F,H)_{B_{s1}  B_s^{\ast} \gamma}$ and 
$G_{B_{s1}  B_{s0}^{\ast} \gamma}$  
are the corresponding effective coupling constants.
The coherent sum of all the diagrams in Figs.4-7, contributing 
to the radiative decays of $B_{s0}^\ast$ and  
$B_{s1}$ mesons, is gauge invariant, while the contribution 
of each diagram is definitely not gauge
invariant. As done in Ref.~\cite{Faessler:2007gv}, for convenience we split 
each individual diagram into a gauge-invariant piece and a remainder, 
which is noninvariant. One can prove that the sum of the noninvariant 
terms vanishes due to gauge invariance. In the following discussion of 
the numerical results we will only deal with the gauge-invariant 
contribution of the separate diagrams of Figs.4-7. In Appendix A we present
the calculational technique for determining the effective couplings 
entering in the matrix elements of the strong and radiative transitions 
of $B_{s0}^\ast$ and $B_{s1}$ mesons. 

Using Eqs.~(\ref{Minv_str}) and (\ref{Minv_em}) the strong 
$B_{s0}^\ast \to B_s \pi^0$, $B_{s1} \to B_s^\ast \pi^0$ 
and radiative $B_{s0}^\ast \to B_s^\ast \gamma$, 
$B_{s1} \to B_s \gamma$, $B_{s1} \to B_s^\ast \gamma$, 
$B_{s1} \to B_{s0}^\ast \gamma$ decay widths are calculated according to 
the expressions: 
\seq 
\eq 
\Gamma(B_{s0}^{\ast} \to B_s \pi^0) \, & = & \,
\frac{G_{B_{s0}^{\ast} B_s \pi}^2}{8\pi m_{B_{s0}^\ast}^2} \, P_{\pi 1} \,, \\
\Gamma(B_{s1} \to B_s^\ast \pi^0) & = & \frac{P_{\pi 2}}{12 \pi
m_{B_{s1}}^2} \, \biggl\{ G_{B_{s1} B_s^\ast \pi}^2 + \frac{1}{2}
\biggl( G_{B_{s1} B_s^\ast \pi} \, w - F_{B_{s1} B_s^\ast \pi}
(w^2 - 1) \biggr)^2 \biggr\} \,,
\en 
\sen 
and 
\seq 
\eq
\Gamma(B_{s0}^{\ast} \to B_s^{\ast} \gamma) & = & \alpha \,
G_{B_{s0}^{\ast} B_s^{\ast} \gamma}^2 \, P_{\gamma 1}^3 \,, \\
\Gamma(B_{s1} \to B_s \gamma) & = & \frac{ \alpha}{3} \, 
G_{B_{s1}B_s \gamma}^2 \, P_{\gamma 2}^{3} \,, \\
\Gamma(B_{s1} \to B_s^\ast \gamma) & = &
\frac{\alpha}{3} \, P_{\gamma 3}^5 \,
\biggl\{ \biggl(G_{B_{s1}  B_s^{\ast} \gamma} 
+ F_{B_{s1}  B_s^{\ast} \gamma}  \biggr)^2 + 
\frac{m_{B_{s1}}^2}{m_{B_s^\ast}^2} \biggl( G_{B_{s1}  B_s^{\ast} \gamma} 
+ H_{B_{s1}  B_s^{\ast} \gamma} \biggr)^2 \biggr\} \,, \\
\Gamma(B_{s1} \to B_{s0}^\ast \gamma) & = & \frac{ \alpha}{3} \, 
G_{B_{s1}B_{s0}^\ast \gamma}^2 \, P_{\gamma 4}^3 \,, 
\en 
\sen 
where $w = v v^\prime = (m_{B_{s1}}^2 + m_{B_s^\ast}^2 
- m_{\pi^0}^2)/(2 m_{B_{s1}} m_{B_s^\ast})$ and 
$P_{\pi i}$, $P_{\gamma i}$ are the corresponding 
three-momenta of the decay products. 

Note, the contribution of the effective 
coupling constant $F_{B_{s1} B_s^\ast \pi}$ to the 
$B_{s1} \to B_s^\ast \pi^0$ decay width is strongly suppressed. 
This is because the contribution of the matrix element with 
$F_{B_{s1} B_s^\ast \pi}$ is proportional to the suppressed factor 
$w^2 - 1 \simeq 4 \times 10^{-3}$ with $w \simeq 1$. Therefore, we have 
\eq 
\Gamma(B_{s1} \to B_s^\ast \pi^0) \ \simeq \ 
\frac{G_{B_{s1} B_s^\ast \pi}^2}{8 \pi m_{B_{s1}}^2} \, P_{\pi 2} \, 
\en  
and 
\eq 
\frac{\Gamma(B_{s1} \to B_s^\ast \pi^0)}
{\Gamma(B_{s0}^{\ast} \to B_s \pi^0)} \ \simeq \  
\frac{P_{\pi 2}}{P_{\pi 1}} \, 
\biggl(\frac{m_{B_{s0}^\ast}}{m_{B_{s1}}}\biggr)^2 \, 
\biggl(\frac{G_{B_{s1}B_s^\ast\pi}}{G_{B_{s0}^\ast B_s\pi}}\biggr)^2 \,. 
\en 

\subsection{Numerical results} 

First, we discuss the choice for the strong coupling constants in 
the Lagrangian ${\cal L}_{\rm str}$~(\ref{L_full}). 
In Refs.~\cite{Faessler:2007gv,Faessler:2007us} we used the set of
strong coupling constants $g_{_{D_1D_2L}}$ $(g_{_{D^\ast D \pi}}$, 
$g_{_{D_s D^\ast K}} = g_{_{D_s K^\ast D}}$ and 
$g_{_{D_s^\ast D K}} = g_{_{D_s^\ast D^\ast K^\ast}})$ 
defined in the charm sector,  
where index $L$ denotes a light meson, while $D_1$ and $D_2$ are the
respective charm states. The coupling $g_{_{D^\ast D \pi}} = 17.9$
was deduced using data for the corresponding strong decay 
width~\cite{Anastassov:2001cw}. The coupling constants 
$g_{_{D_s D^\ast  K}}$ and $g_{_{D_s^\ast D K}}$ have been estimated 
using two different variants of the QCD sum rule approach discussed in 
Refs.~\cite{Wang:2006id,Bracco:2006xf}, where similar results have 
been obtained. Both Refs.~\cite{Wang:2006id,Bracco:2006xf} point to a strong
suppression of these constants  
in comparison to the coupling $g_{_{D^\ast D \pi}}$.
An updated analysis for $g_{_{D^\ast D \pi}}$ in the context of 
a QCD sum rule approach gives a result close to data -- 
$g_{_{D^\ast D \pi}} = 14 \pm 1.5$~\cite{Navarra:2001ju}.  
We used the predictions of  
Ref.~\cite{Wang:2006id}: $g_{_{D^\ast D_s K}} = 2.02$ and 
$g_{_{D_s^\ast D K}} = 1.84$. 
For the unknown parameters $g_{_{D_s K^\ast D}}$ 
and $g_{_{D_s^\ast D^\ast K^\ast}}$ we used the approximative relations 
$g_{_{D_s K^\ast D}} \simeq g_{_{D_s D^\ast K}}$ and 
$g_{_{D_s^\ast D^\ast K^\ast}} \simeq g_{_{D_s^\ast D K}}$, which can 
be explained phenomenologically: the first relation -- 
by the universality of the coupling of the $D_s$ meson to 
$D^\ast K$ and $K^\ast D$ mesonic pairs (it is based on exact SU(4) flavor 
symmetry and we do not expect a substantial violation of this relation 
due to breaking of the SU(4) symmetry) and the second relation -- 
by the universality of the coupling of the $D_s^\ast$ meson to
two pseudoscalars and two vectors (like for $\rho\pi\pi$ and  
$\rho\rho\rho$ couplings: $g_{\rho\pi\pi} \simeq g_{\rho\rho\rho} 
\simeq 6$, and also for $J/\Psi D D$ and 
$J\Psi D^\ast D^\ast$: $g_{J/\Psi D D} \simeq  
g_{J/\Psi D^\ast D^\ast} \simeq 8$~\cite{Lin:1999ad}). For consistency, 
in the present manuscript we use the set of charmed hadronic couplings 
predicted by QCD sum rules (central values) with : 
\eq 
g_{_{D^\ast D \pi}} = 14\,, \hspace*{.5cm}
g_{_{D_s D^\ast K}} = g_{_{D_s K^\ast D}} = 2.02 \,, \hspace*{.5cm}
g_{_{D_s^\ast D K}} = g_{_{D_s^\ast D^\ast K^\ast}} = 1.84  \,. 
\en 
In particular, instead of $g_{_{D^\ast D \pi}} = 17.9$ we use 
$g_{_{D^\ast D \pi}} = 14$. Such a modification does not change the 
numerical results of Refs.~\cite{Faessler:2007gv,Faessler:2007us}, 
because the contribution of the corresponding diagrams containing 
the coupling $g_{_{D^\ast D \pi}}$ is strongly suppressed. 
A similar picture we also have in the bottom sector (see discussion below). 

In order to evaluate the corresponding bottom
couplings  $g_{_{B_1B_2L}}$ we use the arguments of heavy hadron
chiral perturbation theory (HHChPT)~\cite{Wise:1992hn,Casalbuoni:1996pg} 
which relates the bottom and charmed couplings containing the same 
light meson: 
\eq
g_{_{B_1B_2L}} = g_{_{D_1D_2L}} \, \displaystyle\frac{m_B}{m_D}
\en
where $m_B$ and $m_D$ are the masses of the bottom and charm mesons
identified e.g. with the masses of
$m_{B^\pm} = 5.279$~GeV and $m_{D^\pm} = 1.8693$~GeV. In other words,
to get the set of bottom meson couplings used in the strong
Lagrangian (\ref{L_strong}) we rescale the corresponding charm
meson couplings by a factor $m_{B^\pm}/m_{D^\pm} \simeq 2.82$.
In particular, we have:
\eq
g_{_{B^\ast B \pi}}  = 39.5 \,, \hspace*{.5cm}
g_{_{B_s B^\ast K}} \simeq g_{_{B_s K^\ast B}} = 5.70 \,, \hspace*{.5cm}
g_{_{B_s^\ast B K}} \simeq g_{_{B_s^\ast B^\ast K^\ast}} = 5.19  \,.
\en 
For the coupling $g_{B_s^\ast B^\ast K}$ we have no further input. 
From a dimensional analysis it should be of order 1 GeV$^{-1}$. 
The diagrams 6(e) and 6(f), however, where this coupling enters, are 
strongly suppressed, and we therefore do not need to have precise 
knowledge of this constant.  

The coupling $g_{_{K^\ast K \pi}}  = 4.61$ is fixed from 
data on the $K^\ast K \pi$ decay width~\cite{Yao:2006px}.
Finally, we fix the couplings $g_{_{B_{s0}^\ast}}$ and $g_{_{B_{s1}}}$,
which are given by Eqs.~(\ref{coupling_Bs0}) and (\ref{coupling_Bs1})
in terms of the adjustable vertex function. Using the Gaussian vertex
function we obtain the result that these couplings are quite stable with
respect to a variation of the scale parameter $\Lambda_M$.
In particular, varying $\Lambda_M$ from 1 to 2~GeV,
we get a range of values for $g_{_{B_{s0}^\ast}}$ from
27.17 to 23.21~GeV and for $g_{_{B_{s1}}}$ from~25.64 to 22.14~GeV. 
Note that our predictions for these couplings are in agreement with 
the results of other theoretical approaches: 
$g_{_{B_{s0}^\ast}} = g_{_{B_{s1}}} = 19.6 \pm 5.7$~GeV 
[light--cone QCD sum rules approach~\cite{Wang:2008tm}] and 
$g_{_{B_{s0}^\ast}} = 23.572$~GeV and $g_{_{B_{s1}}} = 23.442$~GeV 
[effective chiral approach~\cite{Guo:2006fu,Guo:2006rp}]. 

Now we present the numerical results.  
First, we discuss the results for strong decays. 
Here the main contribution to the decay width comes, 
as expected, from the diagrams of Figs.2(a), (b) and 3(a), (b). 
On the other hand, the contribution of the direct mechanism is comparable 
to the one of the indirect mechanism (i.e. due to $\eta-\pi^0$ mixing). 
The contribution of the diagrams in Figs.2(c) and (d) is of order 0.1\% 
of the total contribution to the $G_{B_{s0}^\ast B_s \pi}$ coupling and 
the contribution of Figs.3(c) and (d) is of order 1.6\% 
of the total contribution to the $G_{B_{s1} B_s^\ast \pi}$ coupling. 
Therefore, the couplings $G_{B_{s0}^\ast B_s \pi}$ and 
$G_{B_{s1} B_s^\ast \pi}$ are not sensitive to a variation of the couplings  
$g_{_{B^\ast B \pi}}$, $g_{_{B_s^\ast B K}}$ and $g_{_{B_s B^\ast K}}$. 
They are only sensitive to the values of the couplings $g_{_{B_s K^\ast B}}$ 
and $g_{_{B_s^\ast B^\ast K^\ast}}$. The leading-order 
contributions from the diagrams in Figs.2(a), (b) and 3(a), (b) 
to the quantities $G_{B_{s0}^\ast B_s \pi}$ and 
$G_{B_{s1} B_s^\ast \pi}$ in terms of the couplings  
$g_{_{B_s K^\ast B}}$ and $g_{_{B_s^\ast B^\ast K^\ast}}$ 
are given by  
\eq 
G_{B_{s0}^\ast B_s \pi} = 65.9 \ g_{_{B_s K^\ast B}} \ {\rm MeV}\,, 
\hspace*{.5cm} 
G_{B_{s1} B_s^\ast \pi} = 72.3 \ g_{_{B_s^\ast B^\ast K^\ast}} \ {\rm MeV} 
\en
for a typical value of the dimensional parameter $\Lambda_M = 1$ GeV.
Then, using the specific values of $g_{_{B_s K^\ast B}} = 5.70$ and 
$g_{_{B_s^\ast B^\ast K^\ast}} = 5.19$ we 
get the following predictions for the effective couplings and decay widths: 
\eq\label{GBs0Bs1}  
G_{B_{s0}^\ast B_s \pi} = 375.7 (375.6) \ {\rm MeV}\,, \hspace*{.5cm} 
G_{B_{s1} B_s^\ast \pi} = 381.1 (376.3) \ {\rm MeV} 
\en 
and 
\eq 
\Gamma(B_{s0}^\ast \to B_s \pi^0) = 55.2 (55.1) \ {\rm keV}\,, \hspace*{.5cm}
\Gamma(B_{s1} \to B_s^\ast \pi^0) = 57.0 (55.6) \ {\rm keV} \,. 
\en 
In the brackets we indicate the results of the leading diagrams of
Figs.2(a), (b) and 3(a), (b). 

The strong decay couplings 
$G_{B_{s0}^\ast B_s \pi}$, $G_{B_{s1} B_s^\ast \pi}$ and the decay 
widths $\Gamma(B_{s0}^{\ast} \to B_s \pi)$,  
$\Gamma(B_{s1} \to B_s^\ast \pi)$ are practically degenerate, which 
can also be explained by heavy quark symmetry (HQS), 
which is a good symmetry  
for the heavy-light mesons with a bottom quark. Because of the 
infinitely heavy mass of the bottom quark the spins of the $\bar b$ 
antiquark and the $s$ quark decouple (spin symmetry), and, therefore, 
the properties of $B_{s0}^\ast$ and $B_{s1}$ mesons become similar. 
This issue was also discussed in Refs.~\cite{Colangelo:2003vg,Mehen:2004uj} 
in the context of the charm partners ($D_{s0}^\ast$ and $D_{s1}$). 
In particular, it was shown that the corresponding coupling constants and 
widths are degenerate in the heavy quark limit. In our previous papers we also 
reproduced the same result. Moreover, for finite charmed meson masses 
the decay characteristics are nearly degenerate: 
\eq\label{GDs0Ds1} 
G_{D_{s0}^\ast D_s \pi} = 145.4 \ {\rm MeV}\,, \hspace*{.5cm} 
G_{D_{s1} D_s^\ast \pi} = 160.2 \ {\rm MeV} 
\en 
and 
\eq 
\Gamma(D_{s0}^\ast \to D_s \pi^0)   = 46.7 \ {\rm keV}\,, \hspace*{.5cm}
\Gamma(D_{s1} \to D_s^\ast \pi^0) = 50.1 \ {\rm keV} \,. 
\en 
The next point is a check of the flavor content of HQS. 
In our molecular approach the leading contributions to the 
couplings $G_{H_{s0}^\ast H_s \pi}$ and $G_{H_{s1} H_s^\ast \pi}$, 
where $H=D$ or $B$, are defined by the pairs of diagrams in Figs.2(a,b) 
and 3(a,b), respectively: 
\seq 
\eq 
G_{H_{s0}^\ast H_s \pi} &=& 
g_{_{H_{s0}^\ast}} \, g_{_{K^\ast K \pi}} \, 
g_{_{H_s H K^\ast}} \, I_{KK^\ast H} \,, \\
G_{H_{s1} H_s^\ast \pi} &=& 
g_{_{H_{s1}}} \, g_{_{K^\ast K \pi}} \, 
g_{_{H_s^\ast H^\ast K^\ast}} \, I_{KK^\ast H^\ast} \,, 
\en 
\sen 
where $I_{KK^\ast H}$ and $I_{KK^\ast H^\ast}$ are the structure 
integrals which are of order $O(1/m_Q)$ in the inverse heavy quark mass 
expansion -- $1/m_Q$. The hadronic couplings scale as: 
\eq 
g_{_{H_{s0}^\ast}}  \sim O(m_Q)\,, \hspace*{.25cm} 
g_{_{H_{s1}}}       \sim O(m_Q)\,, \hspace*{.25cm} 
g_{_{K^\ast K \pi}} \sim O(1)  \,, \hspace*{.25cm} 
g_{_{H_s H K^\ast}} \sim O(m_Q)\,, \hspace*{.25cm} 
g_{_{H_s^\ast H^\ast K^\ast}} \sim O(m_Q)\,. 
\en  
Therefore, the strong couplings 
$G_{H_{s0}^\ast H_s \pi}$ and $G_{H_{s1} H_s^\ast \pi}$  
scale as: 
$G_{H_{s0}^\ast H_s \pi} \sim O(m_Q)$  and 
$G_{H_{s1} H_s^\ast \pi} \sim O(m_Q).$ 
For this scaling behavior the following relations result: 
\eq\label{ratios_fl}  
\frac{G_{B_{s0}^\ast B_s \pi}}{G_{D_{s0}^\ast D_s \pi}} 
\sim \frac{m_b}{m_c}\,, 
\hspace*{.5cm} 
\frac{G_{B_{s1} B_s^\ast \pi}}{G_{D_{s1} D_s^\ast \pi}} 
\sim \frac{m_b}{m_c}\,. 
\en 
With our results [see Eqs.~(\ref{GBs0Bs1}) and (\ref{GDs0Ds1})] 
we conclude that the constraints (\ref{ratios_fl}) are fulfilled very well. 

In Table~\ref{BsS} we present our results for the decay widths 
$\Gamma(B_{s0}^{\ast} \to B_s \pi^0)$ and 
$\Gamma(B_{s1} \to B_s^{\ast} \pi^0)$ including a variation of the 
scale parameter $\Lambda_M$ from $1$ to $2$ GeV (an increase of
$\Lambda_M$ leads to an increase of the widths) and compare them with 
known theoretical predictions~\cite{Bardeen:2003kt,Guo:2006fu,Guo:2006rp,%
Wang:2008ny}.  
Our results for the decay widths are larger in comparison 
to previous approaches~\cite{Bardeen:2003kt,Guo:2006fu,Guo:2006rp,Wang:2008ny} 
due to inclusion of the direct isospin-violating transitions 
$B_{s0}^{\ast} \to B_s \pi^0$ and $B_{s1} \to B_s^\ast \pi^0$.   

Now we turn to the discussion of the radiative decays 
$B_{s0}^{\ast} \to B_s^{\ast} \gamma$ and $B_{s1} \to B_s \gamma$. 
The main contribution to the decay characteristics of 
the $B_{s0}^{\ast} \to B_s^{\ast} \gamma$ and $B_{s1} \to B_s \gamma$
decays comes, 
as expected, from the diagrams of Figs.4(a) and 5(a).  
Our results for the effective coupling constants 
and decay widths for a typical value of $\Lambda_M= 1$ GeV are:  
\eq\label{couplings_bottom}  
G_{B_{s0}^{\ast} B_s^\ast \gamma} \ = \ 0.122 \ {\rm GeV}^{-1}\,, 
\hspace*{.5cm} 
G_{B_{s1} B_s \gamma} \ = \  0.115 \ {\rm GeV}^{-1}
\en 
and 
\eq\label{Gamm_rad} 
\Gamma(B_{s0}^\ast \to B_s^\ast \gamma) \ = \ 3.07 \ {\rm keV}\,, 
\hspace*{.5cm} 
\Gamma(B_{s1} \to B_s \gamma) \ = \ 2.01 \ {\rm keV} \,. 
\en 
In Table~\ref{BsEM} we summarize our results for the radiative decay 
widths including a variation of the scale parameter $\Lambda_M$ from 
1 to 2 GeV (an increase of $\Lambda_{B_M}$ leads to a larger value for 
the width). In comparison, we also display the predictions of other 
theoretical approaches~\cite{Bardeen:2003kt,Vijande:2007ke,Wang:2008bt}. 
The lower limit of the QCD sum rule results~\cite{Wang:2008bt} 
is consistent with our predictions. In our opinion the predictions 
of~\cite{Bardeen:2003kt,Vijande:2007ke} are overestimated. 
In particular, applying HQS (spin symmetry) 
we can relate the corresponding radiative coupling constants of the 
same flavor as  
\eq\label{const_HQL} 
G_{D_{s0}^{\ast} D_s^\ast \gamma} = G_{D_{s1} D_s \gamma}\,, 
\hspace*{.5cm}  
G_{B_{s0}^{\ast} B_s^\ast \gamma} =  G_{B_{s1} B_s \gamma} \,. 
\en  
The same relation was derived previously in HHChPT~\cite{Mehen:2004uj} 
in the charm sector. Now we perform the same exercise  
as done for the strong couplings in order to relate the radiative couplings 
of different flavors. The leading contributions to the 
couplings $G_{H_{s0}^\ast H_s^\ast \gamma}$ and $G_{H_{s1} H_s \gamma}$, 
are defined by the pairs of diagrams in Figs.4(a)  
and 5(a), respectively: 
\seq 
\eq 
G_{H_{s0}^\ast H_s^\ast \gamma} &=& 
g_{_{H_{s0}^\ast}} \, g_{_{K K \gamma}} \, 
g_{_{H_s^\ast H K}} \, I_{KKH} \,, \\
G_{H_{s1} H_s \gamma} &=& 
g_{_{H_{s1}}} \, g_{_{K K \gamma}} \, 
g_{_{H_s H^\ast K}} \, I_{KKH^\ast} \,, 
\en  
\sen 
where $I_{KKH}$ and $I_{KKH^\ast}$ are the structure 
integrals which are of order $O(1/m_Q^2)$. 
The couplings $g_{_{K K \gamma}}$, $g_{_{H_s^\ast H K}}$ 
and $g_{_{H_s H^\ast K}}$ scale as:   
$g_{_{K K \gamma}}   \sim O(1)\,,$   
$g_{_{H_s^\ast H K}} \sim O(m_Q)$ and  
$g_{_{H_s  H^\ast K}} \sim O(m_Q).$  
Therefore, the radiative couplings  
$G_{H_{s0}^\ast H_s^\ast \gamma}$ and $G_{H_{s1} H_s \gamma}$  
are insensitive to the flavor of the heavy quark/meson:  
$G_{H_{s0}^\ast H_s \gamma} \sim O(1)$ and 
$G_{H_{s1} H \gamma} \sim O(1).$ 
Finally, the radiative couplings obey both spin and 
flavor symmetry in the heavy quark limit and we arrive at the 
constraint: 
\eq\label{ratios_rad_full}  
G_{D_{s0}^{\ast} D_s^\ast \gamma} = G_{D_{s1} D_s \gamma} = 
G_{B_{s0}^{\ast} B_s^\ast \gamma} = G_{B_{s1} B_s \gamma} \,. 
\en 
Recalling the results for the radiative 
decay constants of charmed mesons~\cite{Faessler:2007gv,Faessler:2007us} 
\eq\label{couplings_charm} 
G_{D_{s0}^{\ast} D_s^\ast \gamma} \ = \ 0.093 \ {\rm GeV}^{-1}\,, 
\hspace*{.5cm} 
G_{D_{s1} D_s \gamma} \ = \  0.106 \ {\rm GeV}^{-1}
\en 
we conclude that relation (\ref{ratios_rad_full}) is approximately fulfilled 
by our results, which are obtained for finite physical masses of the 
heavy mesons. The constraint (\ref{ratios_rad_full}) 
can also be used to deduce relations between the corresponding decay 
widths:  
\seq 
\eq
& &\frac{\Gamma(D_{s1} \to D_s \gamma)} 
        {\Gamma(D_{s0}^\ast \to D_s^\ast \gamma)}
= \frac{1}{3} \, 
\biggl(\frac{m_{D_{s1}}^2 - m_{D_s}^2} 
        {m_{D_{s0}^\ast}^2 - m_{D_s^\ast}^2}\biggr)^3
\, \biggl(\frac{m_{D_{s0}^\ast}}{m_{D_{s1}}}\biggr)^3  \simeq 3.80 \,, 
\label{Ds10}\\ 
& &\frac{\Gamma(B_{s1} \to B_s \gamma)} 
        {\Gamma(B_{s0}^\ast \to B_s^\ast \gamma)}
= \frac{1}{3} \, \biggl(\frac{m_{B_{s1}}^2 - m_{B_s}^2} 
        {m_{B_{s0}^\ast}^2 - m_{B_s^\ast}^2}\biggr)^3
\, \biggl(\frac{m_{B_{s0}^\ast}}{m_{B_{s1}}}\biggr)^3  \simeq 0.74 \,, 
\label{Bs10}
\en 
and
\eq
& &\frac{\Gamma(B_{s0}^\ast \to B_s^\ast \gamma)}
        {\Gamma(D_{s0}^\ast \to D_s^\ast \gamma)}
= \biggl(\frac{m_{B_{s0}^\ast}^2 - m_{B_s^\ast}^2}
        {m_{D_{s0}^\ast}^2 - m_{D_s^\ast}^2}\biggr)^3
\, \biggl(\frac{m_{D_{s0}^\ast}}{m_{B_{s0}^\ast}}\biggr)^3  \simeq 3.74 \,,
\label{BDs0}\\
& &\frac{\Gamma(B_{s1} \to B_s \gamma)}{\Gamma(D_{s1} \to D_s \gamma)}
= \biggl(\frac{m_{B_{s1}}^2 - m_{B_s}^2}{m_{D_{s1}}^2 - m_{D_s}^2}\biggr)^3
\, \biggl(\frac{m_{D_{s1}}}{m_{B_{s1}}}\biggr)^3 \simeq  0.73 \,. 
\label{BDs1} 
\en 
\sen 
Relation (\ref{Ds10}) is confirmed by the full analysis done 
in our framework and in other theoretical approaches (see compilation 
of the results in Refs.~\cite{Faessler:2007gv,Faessler:2007us}). 
E.g. the relation explains why most of the approaches predict that 
the decay width $\Gamma(D_{s1} \to D_s \gamma)$ is approximately 
$3 - 5$ times larger than $\Gamma(D_{s0}^\ast \to D_s^\ast \gamma)$. 
The other relations (\ref{Bs10})-(\ref{BDs1}) 
help to give predictions for the decay widths of the bottom partners. 
In particular, we arrive at the conclusion that our full predictions 
for $\Gamma(B_{s0}^\ast \to B_s^\ast \gamma)$ and 
$\Gamma(B_{s1} \to B_s \gamma)$ -- a few keV -- are well justified. 

Now we discuss the ratios of radiative and strong decay modes. 
For both systems, $B_{s0}^\ast$ and $B_{s1}$, we predict small ratios: 
\seq  
\eq 
R_{B_{s0}^\ast} &=& \frac{B_{s0}^{\ast} \to B_s^{\ast} \gamma}
{B_{s0}^{\ast} \to B_s \pi} \simeq 0.05\,, \nonumber\\
R_{B_{s1}} &=& \frac{B_{s1} \to B_s \gamma}
{B_{s1} \to B_s^{\ast} \pi} \simeq 0.03 \,. 
\en 
\sen 
Note that similar results we also obtained in the charm sector: 
$R_{D_{s0}^\ast} = 0.01$ and $R_{D_{s1}} = 0.05$. 
The predicted ratio $R_{D_{s0}^\ast}$ is consistent with the present 
experimental limit $R_{D_{s0}^\ast} < 0.059$, while the ratio $R_{D_{s1}}$ 
is smaller than the result quoted by the Particle Data Group 
$R_{D_{s1}} =  0.38 \pm 0.05$~\cite{Yao:2006px}. 
We consider this experimental result 
as preliminary, since it is not clear why the ratio for the axial 
state $D_{s1}$ is much larger than for the scalar state $D_{s0}^\ast$. 
At this point more precise data on the strong and radiative 
decays of the $D_{s1}$ meson and their ratio would be very helpful. 

Finally, we give the predictions for the other two radiative decay 
modes of $B_{s1}$ mesons --  
$B_{s1} \to B_s^{\ast} \gamma$ and $B_{s1} \to B_{s0}^{\ast} \gamma$. 
These decay amplitudes are generated by the anomalous couplings 
of two vectors and one pseudoscalar and one can expect that they 
are suppressed. Moreover, there is an additional mechanism for their 
suppression. The leading diagrams in the 
process $B_{s1} \to B_s^{\ast} \gamma$ are the ones of Figs.6(a) and (b). 
The separate contribution of the diagrams in Figs.6(c)-(f) to the 
corresponding decay width is of order 1\%. The diagrams of Figs.6(a) 
and Figs.6(b) are subtracted from each other 
because of the opposite sign of the anomalous couplings 
$g_{_{K^{\ast \, \pm} \to K^\pm \gamma}} = 0.836$ GeV$^{-1}$ 
and 
$g_{_{K^{\ast \, 0} \to K^0 \gamma}} = -1.267$ GeV$^{-1}$. 
In the case of the 
$B_{s1} \to B_{s0}^{\ast} \gamma$ decay we only have two diagrams 
contributing to the matrix element. Again, due to the opposite sign of 
the anomalous couplings 
$g_{_{B^{\ast \, \pm} \to B^\pm \gamma}} = 0.5$ GeV$^{-1}$ 
and $g_{_{B^{\ast \, 0} \to B^0 \gamma}} = -1$ GeV$^{-1}$ their total 
contribution is given by the difference of the individual contributions. 
Finally, as a full result we get values of 0.04 to 0.18 keV for
the decay width of $B_{s1} \to B_s^{\ast} \gamma$ 
including a variation
of the model parameter $\Lambda$ from 1 to 2 GeV. The result for the 
decay width of $B_{s1} \to B_{s0}^{\ast} \gamma$ is stable with respect 
to a variation of  the model parameter $\Lambda$ in the same range and 
is equal to 0.022 keV. 
We also present the result for $\Gamma(B_{s1} \to B_{s0}^{\ast} \gamma)$
in terms of the couplings $g_{_{B^{\ast \, \pm} \to B^\pm \gamma}}$ 
and and $g_{_{B^{\ast \, 0} \to B^0 \gamma}}$: 
\eq 
\Gamma(B_{s1} \to B_{s0}^{\ast} \gamma) = 
[ (g_{_{B^{\ast \, \pm} \to B^\pm \gamma}} 
+ g_{_{B^{\ast \, 0} \to B^0 \gamma}}) \times 1 \ {\rm GeV}]^2 
\ 0.088 \ {\rm keV} \,. 
\en 
Recently, these radiative decay widths have also been 
estimated using light-cone QCD sum rules~\cite{Wang:2008bt}: 
$\Gamma(B_{s1} \to B_s^\ast \gamma) = 0.3 - 6.1$ keV and 
$\Gamma(B_{s1} \to B_{s0}^{\ast} \gamma) = 0.002 - 0.008$ keV. 
The decay width
$\Gamma(B_{s1} \to B_{s0}^{\ast} \gamma)$ is definitely strongly supressed 
in both approaches. The decay width $\Gamma(B_{s1} \to B_s^\ast \gamma)$ 
predicted in~\cite{Wang:2008bt} is also smaller in comparison to the other 
two modes $B_{s0}^\ast \to B_s^\ast \gamma$ and $B_{s1} \to B_s \gamma$
(see Table~\ref{BsEM}).  

\newpage 

\section{Summary}

In this paper we studied the new bottom-strange mesons
$B_{s0}^\ast(5725)$ and $B_{s1}(5778)$ in the hadronic molecule
interpretation, i.e., we considered them as bound states of $BK$ and 
$B^\ast K$ mesons using a phenomenological Lagrangian approach.  
Our approach is based on the compositeness condition 
with $Z=0$ (the renormalization constant of the molecular state equals zero). 
This condition is crucial for weakly bound states. The compositeness
condition is a self-consistent tool to evaluate for a hadronic molecule,
once the mass of the bound state is fixed, its coupling to the intermediate
hadronic state, which in turn feeds the accessible final states.
Once the further hadronic couplings to generate the final states
are known, this approach is fairly model-independent in the results
for the observable decay modes.
Furthermore, the Lagrangian method implied by the compositeness
condition allows a fully Lorentz and gauge-invariant treatment of
the problem. 

We calculated the strong $B_{s0}^{\ast} \to B_s \pi^0$, 
$B_{s1} \to B_s^{\ast} \pi^0$ and radiative 
$B_{s0}^{\ast} \to B_s^{\ast} \gamma$, $B_{s1} \to B_s \gamma$, 
$B_{s1} \to B_s^{\ast} \gamma$, $B_{s1} \to B_{s0}^{\ast} \gamma$ 
decays. A new impact of the $B K$ and $B^\ast K$ 
molecular structures of the $B_{s0}^\ast(5725)$ and
$B_{s1}(5778)$ mesons is that the presence of $u(d)$ quarks in the
$B^{(\ast)}$ and $K$ meson loops gives rise to direct strong
isospin-violating transitions $B_{s0}^{\ast} \to B_s \pi^0$ and
$B_{s1} \to B_s^{\ast} \pi^0$ in addition to the decay mechanism
induced by $\eta-\pi^0$ mixing. We showed that the direct transition 
is comparable with the $\eta-\pi^0$ mixing transition. As a consequence, 
the presence of the direct mode makes our predictions larger than
the ones of previous approaches.
In the case of the radiative decays $B_{s0}^{\ast} \to B_s^{\ast}
\gamma$ and $B_{s1} \to B_s \gamma$, our results are considerably smaller 
than in previous calculations. We also gave predictions for 
the anomalous decays $B_{s1} \to B_s^{\ast} \gamma$ and 
$B_{s1} \to B_{s0}^{\ast} \gamma$: their decay widths are 
suppressed in comparison to the leading radiative modes 
$B_{s0}^{\ast} \to B_s^{\ast} \gamma$ and $B_{s1} \to B_s \gamma$. 

In series of papers~\cite{Faessler:2007gv}-\cite{Faessler:2007us} 
we already studied in detail strong, electromagnetic and weak decays 
of $D_{s0}^\ast$ and $D_{s1}$ states, as a consequence 
of their possible molecular structure. At present we do not think that 
the structure issue concerning the $D_{s0}^\ast$ and $D_{s1}$ mesons 
is settled yet, in particular because the mass and the narrowness of these 
states cannot be easily explained in the context of the standard $c \bar s$ 
picture~\cite{Rosner:2006vc,Barnes:2003dj}. 
A further direct consequence and analogy of possible hadronic molecules
in the $D_s$ sector is also found in the $B_s$ system. If the $B_{s0}^\ast$ 
and $B_{s1}$ mesons can possibly be experimentally established
near the predicted mass values, then, because of their closeness to the
$BK$ and $BK^\ast$ thresholds, they are clear candidates for hadronic 
molecules. In the present approach we give clear predictions for the possible
strong and radiative decay modes of $B_{s0}^\ast$ and $B_{s1}$ mesons 
accessible by experiment. Strong experimental deviations from our results 
would discard the molecular interpretation of these states.        
We hope that our results will be useful for future experiments, 
where $B_{s0}^{\ast}$ and $B_{s1}$ could possibly be detected. 

\acknowledgments 

This work was supported by the DFG under Contract Nos. FA67/31-1 and
GRK683. This research is also part of the EU
Integrated Infrastructure Initiative Hadronphysics project under 
Contract No. RII3-CT-2004-506078 and the President Grant of Russia
``Scientific Schools''  No. 871.2008.2 

\newpage 

\appendix\section{Effective couplings for
strong and radiative transitions of $B_{s0}^\ast$ and $B_{s1}$ mesons.}

Here we discuss the calculational technique of the matrix elements of 
strong and radiative transitions of $B_{s0}^\ast$ and $B_{s1}$ mesons. 

As an example for the calculation of diagrams presented in Figs.1-7 
we consider a generic loop integral containing 
a product of $n$ virtual momenta $k_{\mu_1} \cdots 
k_{\mu_n}$, three meson propagators with masses $m_1$, $m_2$ and $m_3$ 
and the correlation function of the molecular state 
($B_{s0}^\ast$ or $B_{s1}$ meson)
\eq 
I_{\mu_1 \cdots \mu_n}(p,p^\prime) = 
\int\frac{d^4 k}{\pi^2 i} \tilde\Phi(-k^2) \ 
\frac{k_{\mu_1} \cdots 
k_{\mu_n}}{(m_1^2 - (k+p)^2) (m_2^2 - (k+p^\prime)^2) (m_3^2 - k^2)}\; . 
\en 
The three main ingredients are
\begin{itemize}
\item  use of the Laplace transform of the vertex function 
$$
\tilde \Phi(-z)=\int\limits_0^\infty\! ds\, \tilde \Phi_L(s)\, e^{sz}\;, 
$$ 
which is useful to proceed with vertex functions of any functional form, 
\vspace{-0.2cm}
\item the  $\alpha$-transform of the denominator
$$
\frac{1}{m_i^2-k_i^2}=\int\limits_0^\infty\! d\alpha \,  
e^{-\alpha (m_i^2-k_i^2)}\; , 
$$
\vspace{-0.2cm}
\item the differential representation of the numerator
$$
k_{\mu_i} \, e^{kR}=
\frac{\partial}{\partial R^{\mu_i}} e^{kR}\; ,
$$ 
where $R$ is a linear combination of the external momenta. 
\end{itemize} 
The calculation  of the transition form factors amounts to a one-loop
integration. Integration over the loop momentum is done analytically.
One ends up with the integrals over $\alpha$ Feynman parameters 
which are not difficult to evaluate 
numerically. All calculations are done by using computer programs written 
in FORM~\cite{Vermaseren:2000nd} and in FORTRAN for numerical 
evaluations. 
Also, for transparency we give explicit expressions for the leading 
contributions to the effective couplings defining the structure 
of the matrix elements of strong and radiative decays of 
$B_{s0}^\ast$ and $B_{s1}$ mesons. 

\subsection{Decay $B_{s0}^\ast \to B_s \pi^0$.}

The leading contribution to the coupling constant $G_{B_{s0}^\ast B_s \pi}$ 
coming from the diagrams in Figs.2(a) and 2(b) is defined by 
\eq 
G_{B_{s0}^\ast B_s \pi} = \frac{1}{16 \pi^2} \
g_{B_{s0}^\ast} \ g_{K^\ast K \pi} \ g_{B_s B K^\ast} \ 
I_{B_{s0}^\ast B_s \pi} \,. 
\en 
Here $I_{B_{s0}^\ast B_s \pi}$ is the structure integral 
\eq 
I_{B_{s0}^\ast B_s \pi} &=& \sum\limits_{i=1}^5 I_{B_{s0}^\ast B_s \pi}^i 
\,, \nonumber\\
I_{B_{s0}^\ast B_s \pi}^1 &=& 
\int\limits_0^\infty\int\limits_0^\infty\int\limits_0^\infty 
\frac{d\alpha_1 d\alpha_2 d\alpha_3}{\Delta_1^2} \, 
\exp(A_1) \,  N_1 \,, \nonumber\\ 
I_{B_{s0}^\ast B_s \pi}^2 &=& 
\int\limits_0^\infty\int\limits_0^\infty\int\limits_0^\infty 
\frac{d\alpha_1 d\alpha_2}{\Delta_2^2} \, 
\exp(A_2) \,  N_2 \,, \nonumber\\
I_{B_{s0}^\ast B_s \pi}^3 &=& 
\int\limits_0^\infty\int\limits_0^\infty\int\limits_0^\infty 
\frac{d\alpha_1 d\alpha_2}{\Delta_3^2} \, 
\exp(A_3) \,  N_3 \,, \\
I_{B_{s0}^\ast B_s \pi}^4 &=& 
\int\limits_0^\infty\int\limits_0^\infty\int\limits_0^\infty 
\frac{d\alpha_1 d\alpha_2}{\Delta_4^2} \, 
\exp(A_4) \,  N_4 \,, \nonumber\\
I_{B_{s0}^\ast B_s \pi}^5 &=& 
\int\limits_0^\infty\int\limits_0^\infty 
\frac{d\alpha_1}{\Delta_5^2} \, 
\exp(A_5) \,  N_5 \,, \nonumber 
\en 
where we introduce the set of notations: 
\eq 
\Delta_1 = 1 + \alpha_{123}\,, \hspace*{.25cm}   
\Delta_2 = \Delta_3 = \Delta_4 = 1 + \alpha_{12}\,, \hspace*{.25cm}   
\Delta_5 = 1 + \alpha_1\,, \nonumber 
\en 
\eq 
A_1 &=& - \alpha_1 \mu_B^2 - \alpha_2 \mu_{K^{\ast}}^2 - \alpha_3 \mu_K^2 
- \mu_{B_{s0}^{\ast}}^2 w_{BK} w_{KB} + \frac{1}{\Delta_1} 
\biggl(\mu_{B_{s0}^{\ast}}^2 \alpha_{1KB} \alpha_{3BK} 
+ \mu_{B_s}^2 \alpha_{1KB} \alpha_2 
+ \mu_\pi^2 \alpha_{3BK} \alpha_2 \biggr) \,,\nonumber\\ 
A_2 &=& - \alpha_1 \mu_B^2 - \alpha_2 \mu_{K^{\ast}}^2 
- \mu_{B_{s0}^{\ast}}^2 w_{BK} w_{KB} + \frac{1}{\Delta_2} 
\biggl(\mu_{B_{s0}^{\ast}}^2 \alpha_{1KB} w_{BK} 
+ \mu_{B_s}^2 \alpha_{1KB} \alpha_2 
+ \mu_\pi^2 w_{BK} \alpha_2 \biggr) \,,\nonumber\\
A_3 &=& - \alpha_1 \mu_{K^{\ast}}^2 - \alpha_2 \mu_K^2 
- \mu_{B_{s0}^{\ast}}^2 w_{BK} w_{KB} + \frac{1}{\Delta_3} 
\biggl(\mu_{B_{s0}^{\ast}}^2 \alpha_{2BK} w_{KB}  
+ \mu_{B_s}^2 w_{KB} \alpha_1 
+ \mu_\pi^2 \alpha_{2BK} \alpha_1 \biggr) \,, \nonumber \\
A_4 &=& - \alpha_1 \mu_B^2 - \alpha_2 \mu_K^2 
- \mu_{B_{s0}^{\ast}}^2 w_{BK} w_{KB} + \frac{1}{\Delta_4} 
\mu_{B_{s0}^{\ast}}^2 \alpha_{1KB} \alpha_{2BK}  \,,\nonumber\\
A_5 &=& - \alpha_1 \mu_{K^\ast}^2 
- \mu_{B_{s0}^{\ast}}^2 w_{BK} w_{KB} + \frac{1}{\Delta_5} 
\biggl( \mu_{B_{s0}^{\ast}}^2 w_{BK} w_{KB}   
+ \mu_{B_s}^2 w_{KB} \alpha_1 + \mu_\pi^2 w_{BK}\alpha_1 \biggr) 
\,,\nonumber 
\en 
\eq 
N_1 &=& \mu_B^2 + \mu_K - 2 \mu_{B_{s0}^{\ast}}^2 + \mu_{B_s}^2 
+  \mu_\pi^2 -  \mu_{K^{\ast}}^2 
+ \frac{\mu_K^2 - \mu_\pi^2}{\mu_{K^{\ast}}^2} 
(\mu_{B_s}^2 - \mu_B^2) \,,\nonumber\\ 
N_2 &=& 1 + \frac{\mu_{B_s}^2 - \mu_B^2}{\mu_{K^{\ast}}^2} \,,\nonumber\\ 
N_3 &=& - 1 + \frac{\mu_{K}^2 - \mu_\pi^2}{\mu_{K^{\ast}}^2} \,,
\nonumber\\ 
N_4 &=& 1\,, \nonumber\\
N_5 &=& - \frac{1}{\mu_{K^{\ast}}^2}\,,  \nonumber
\en 
\eq 
w_{M_1M_2} = \frac{m_{M_1}  m_{M_2}}{m_{M_1} + m_{M_2}}\,, \hspace*{.25cm} 
\alpha_{123} = \alpha_1 + \alpha_2 + \alpha_3\,, \hspace*{.25cm} 
\alpha_{12} = \alpha_1 + \alpha_2\,, \hspace*{.25cm} 
\alpha_{iM_1M_2} = \alpha_i + w_{M_1M_2}\,, \hspace*{.25cm} 
\mu_M = \frac{m_M}{\Lambda}\,.  \nonumber 
\en 

\subsection{Decay $B_{s1} \to B_s^\ast \pi^0$.}

The leading contributions to the coupling constants 
$G_{B_{s1}^\ast B_s^\ast \pi}$ and 
$F_{B_{s1}^\ast B_s^\ast \pi}$
coming from the diagrams in Figs.3(a) and 3(b) are defined by 
\eq 
G_{B_{s1}^\ast B_s^\ast \pi} 
= \frac{1}{16 \pi^2} \
g_{B_{s1}} g_{K^\ast K \pi} g_{B_s^\ast B^\ast K^\ast} 
I^G_{B_{s1} B_s^\ast \pi} 
\en 
and 
\eq 
F_{B_{s1}^\ast B_s^\ast \pi} 
= \frac{1}{16 \pi^2} \
g_{B_{s1}} g_{K^\ast K \pi} g_{B_s^\ast B^\ast K^\ast} 
I^F_{B_{s1} B_s^\ast \pi} \,. 
\en 
Here $I^G_{B_{s1} B_s^\ast \pi}$ and $I^F_{B_{s1} B_s^\ast \pi}$ 
are the structure integrals 
\eq 
I^G_{B_{s1} B_s^\ast \pi} 
&=& \sum\limits_{i=1}^5 I^{G,i}_{B_{s1} B_s^\ast \pi} \,, 
\nonumber\\
I^{G,1}_{B_{s1} B_s^\ast \pi} &=& 
\int\limits_0^\infty\int\limits_0^\infty\int\limits_0^\infty 
\frac{d\alpha_1 d\alpha_2 d\alpha_3}{\Delta_1^2} \, 
\exp(A_1^G) \,  N_1^G \,, \nonumber\\ 
I^{G,2}_{B_{s1} B_s^\ast \pi} &=& 
\int\limits_0^\infty\int\limits_0^\infty\int\limits_0^\infty 
\frac{d\alpha_1 d\alpha_2}{\Delta_2^2} \, 
\exp(A_2^G) \,  N_2^G \,, \nonumber\\
I^{G,3}_{B_{s1} B_s^\ast \pi} &=& 
\int\limits_0^\infty\int\limits_0^\infty\int\limits_0^\infty 
\frac{d\alpha_1 d\alpha_2}{\Delta_3^2} \, 
\exp(A_3^G) \,  N_3^G \,, \\ 
I^{G,4}_{B_{s1} B_s^\ast \pi} &=& 
\int\limits_0^\infty\int\limits_0^\infty\int\limits_0^\infty 
\frac{d\alpha_1 d\alpha_2}{\Delta_4^2} \, 
\exp(A_4^G) \,  N_4^G \,, \nonumber\\
I^{G,4}_{B_{s1} B_s^\ast \pi} &=& 
\int\limits_0^\infty\int\limits_0^\infty 
\frac{d\alpha_1}{\Delta_5^2} \, 
\exp(A_5^G) \,  N_5^G \,, \nonumber 
\en 
and 
\eq 
I^F_{B_{s1} B_s^\ast \pi} 
&=& \sum\limits_{i=1}^2 I^{F.i}_{B_{s1} B_s^\ast \pi} \,, \nonumber\\
I^{F,1}_{B_{s1} B_s^\ast \pi} &=& 
\int\limits_0^\infty\int\limits_0^\infty\int\limits_0^\infty 
\frac{d\alpha_1 d\alpha_2 d\alpha_3}{\Delta_1^2} \, 
\exp(A_1^F) \,  N_1^F \,, \\ 
I^{F,2}_{B_{s1} B_s^\ast \pi} &=& 
\int\limits_0^\infty\int\limits_0^\infty\int\limits_0^\infty 
\frac{d\alpha_1 d\alpha_2}{\Delta_2^2} \, 
\exp(A_2^F) \,  N_2^F \,, \nonumber 
\en 
where we introduce the set of notations: 
\eq 
A_1^G &=& A_1^F = - \alpha_1 \mu_{K^{\ast}}^2 - \alpha_2 \mu_K^2 
- \alpha_3 \mu_{B^\ast}^2 
- \mu_{B_{s1}}^2 w_{B^\ast K} w_{KB^\ast} \frac{1}{\Delta_1} 
\biggl(\mu_{D_{s1}}^2 \alpha_{1KB^\ast} \alpha_{3B^\ast K} 
+ \mu_{B_s}^2 \alpha_{1KB^\ast} \alpha_2 
+ \mu_\pi^2 \alpha_{3B^\ast K} \alpha_2 \biggr) \,,\nonumber\\ 
A_2^G &=& A_2^F = - \alpha_1 \mu_B^2 - \alpha_2 \mu_{K^{\ast}}^2 
- \mu_{B_{s1}}^2 w_{B^\ast K} w_{KB^\ast } + \frac{1}{\Delta_2} 
\biggl(\mu_{B_{s1}}^2 \alpha_{1KB^\ast } w_{B^\ast K} 
+ \mu_{B_s}^2 \alpha_{1KB^\ast} \alpha_2 
+ \mu_\pi^2 w_{B^\ast K} \alpha_2 \biggr) \,,\nonumber\\
A_3^G &=& - \alpha_1 \mu_{K^{\ast}}^2 - \alpha_1 \mu_K^2 
- \mu_{B_{s1}}^2 w_{B^\ast K} w_{KB^\ast } + \frac{1}{\Delta_3} 
\biggl(\mu_{B_{s1}}^2 \alpha_{2B^\ast K} w_{KB^\ast }  
+ \mu_{B_s}^2 w_{KB^\ast} \alpha_1 
+ \mu_\pi^2 \alpha_{2B^\ast K} \alpha_1 \biggr) \,, \nonumber\\
A_4^G &=& - \alpha_1 \mu_B^2 - \alpha_1 \mu_K^2 
- \mu_{B_{s1}}^2 w_{B^\ast K} w_{KB^\ast} + \frac{1}{\Delta_4} 
\mu_{B_{s1}}^2 \alpha_{1KB^\ast} \alpha_{2B^\ast K}  \,,\nonumber\\
A_5^G &=& - \alpha_1 \mu_{K^\ast}^2 
- \mu_{B_{s1}}^2 w_{B^\ast K} w_{KB^\ast} + \frac{1}{\Delta_5} 
\biggl( 
\mu_{B_{s1}}^2 w_{B^\ast K} w_{KB^\ast} 
+ \mu_{B_s}^2 w_{KB^\ast } \alpha_1 + \mu_\pi^2 w_{B^\ast K}\alpha_1 \biggr) 
\,,\nonumber 
\en 
\eq 
N_1^G &=& 2 \mu_{B_{s1}}^2 - \mu_{B_s^\ast}^2 - \mu_{B^\ast}^2 
+ \mu_{K^{\ast}}^2 - \mu_K^2 + \frac{\mu_K^2}{\mu_{K^{\ast}}^2} 
(  \mu_{B^\ast}^2 - \mu_{B_s^\ast}^2 ) - 
\frac{\mu_{K^{\ast}}^2 - \mu_K^2 }
{ 2 \mu_{B^\ast}^2 \mu_{K^{\ast}}^2 \Delta_1} 
(\mu_{B^\ast}^2 +  \mu_{K^{\ast}}^2 - \mu_{B_s^\ast}^2) \,, 
\nonumber\\ 
N_2^G &=& \frac{\mu_{B_s^\ast}^2 +  \mu_{K^{\ast}}^2 - \mu_{B^\ast}^2} 
{ \mu_{K^\ast}^2}  
- \frac{\mu_{B^\ast}^2 +  \mu_{K^{\ast}}^2 - \mu_{B_s^\ast}^2} 
{2 \mu_{B^\ast}^2 \mu_{K^{\ast}}^2 \Delta_2} \,,\nonumber\\ 
N_3^G &=& - 1 + \frac{1}{2 \mu_{B^{\ast}}^2 \Delta_3} \,, \nonumber\\ 
N_4^G &=& \frac{ \mu_{K^{\ast}}^2 -  \mu_K^2 }{\mu_{K^{\ast}}^2 }\,, 
\nonumber\\
N_5^G &=& \frac{1}{\mu_{K^{\ast}}^2}\,,  \nonumber\\
N_1^F &=& \frac{\mu_{B_{s1}} \mu_{B_s^\ast}}{\Delta_1} 
\biggl( 4 (\alpha_{13} + w_{KB^\ast} ) 
+ \frac{\alpha_3}{\Delta_1} \frac{\alpha_{13} + w_{KB^\ast}} 
{\mu_{B^\ast}^2  \mu_{K^\ast}^2} 
(\mu_{B^\ast}^2 + \mu_{K^\ast}^2 - \mu_{B^\ast_s}^2) 
(\mu_{K^\ast}^2 - \mu_K^2)\nonumber\\
&-& \alpha_3 \frac{(\mu_{K^\ast}^2 + \mu_K^2)
(\mu_{B^\ast_s}^2 - \mu_{K^\ast}^2) + \mu_{B^\ast}^2 
(3 \mu_{K^\ast}^2 - \mu_K^2) } 
{\mu_{B^{\ast}}^2 \mu_{K^{\ast}}^2} \biggr) 
\nonumber\\ 
N_2^F &=& - \frac{\alpha_3}{\Delta_2}  \ 
\frac{\mu_{B_{s1}} \mu_{B_s^\ast}}{\mu_{B^\ast}^2 \mu_{K^{\ast}}^2}  
(\mu_{B^\ast}^2 + \mu_{K^\ast}^2 - \mu_{B_s^\ast}^2)
\biggl( 1 - \frac{\alpha_{13} + w_{KB^\ast}} 
{\Delta_2}  \biggr) \,. \nonumber
\en 

\subsection{Decay $B_{s0}^\ast \to B_s^\ast \gamma$.}

The leading contribution to the coupling constant 
$G_{B_{s0} B_s^\ast \gamma}$
comes from the diagram in Fig.4(a) and is defined by 
\eq
G_{B_{s0}^{\ast} B_s^{\ast} \gamma} = \frac{1}{16 \pi^2 \, \Lambda^2} \,
g_{_{B_{s0}^\ast}} g_{_{B_s^\ast BK}} \, 
I_{B_{s0}^{\ast} B_s^{\ast} \gamma} \,,
\en
where $I_{B_{s0}^{\ast} B_s^{\ast} \gamma}$ is the structure integral 
\eq
I_{B_{s0}^{\ast} B_s^{\ast} \gamma} =  
\int\limits_0^\infty\int\limits_0^\infty\int\limits_0^\infty 
\frac{d\alpha_1 d\alpha_2 d\alpha_3}{\Delta_1^4} \, \exp(B_1) 
\ \alpha_{1BK} \ \alpha_{3KB} \,, 
\en
where
\eq
B_1 = - \alpha_{12} \mu_K^2 - \alpha_3 \mu_B^2  
- \mu_{B_{s0}^{\ast}}^2 w_{BK} w_{KB} + \frac{\alpha_{3KB}}{\Delta_1} 
(\mu_{B_{s0}^{\ast}}^2 \alpha_{1BK} + \mu_{B_s^\ast}^2 \alpha_2 ) \,. 
\en

\subsection{Decay $B_{s1}^\ast \to B_s \gamma$.}

The leading contribution to the coupling constant 
$G_{B_{s1} B_s \gamma}$ is due to
the diagram in Fig.5(a) and is defined by 
\eq
G_{B_{s1} B_s \gamma} = \frac{1}{16 \pi^2 \, \Lambda^2} \,
g_{_{B_{s1}}} g_{_{B_s B^\ast K}} \, 
I_{B_{s1} B_s \gamma} \,,
\en
where $I_{B_{s1} B_s \gamma}$ is the structure integral 
\eq
I_{B_{s1} B_s \gamma} =  
\int\limits_0^\infty\int\limits_0^\infty\int\limits_0^\infty 
\frac{d\alpha_1 d\alpha_2 d\alpha_3}{\Delta_1^3} \, \exp(B_2) \ 
\alpha_{3KB^\ast} \ \biggl( 4 - \frac{2\alpha_1}{\Delta_1} 
\biggl(1 + \frac{\mu_{D_s}^2 - \mu_K^2}{\mu_{D_s^\ast}^2} \biggr) 
\biggr) 
\,, 
\en
where
\eq
B_2 = - \alpha_{12} \mu_K^2 - \alpha_3 \mu_{B^\ast}^2 
- \mu_{B_{s1}}^2 w_{B^\ast K} w_{KB^\ast } 
+ \frac{\alpha_{3KB^\ast}}{\Delta_1} 
(\mu_{B_{s0}^{\ast}}^2 \alpha_{1B^\ast K} + \mu_{B_s}^2 \alpha_2 ) \,. 
\en

\subsection{Decay $B_{s1}^\ast \to B_s^\ast \gamma$.}

The leading contributions to the coupling constant 
$G_{B_{s1} B_s^\ast \gamma}$, 
$F_{B_{s1} B_s^\ast \gamma}$ and
$H_{B_{s1} B_s^\ast \gamma}$ coming from the diagrams 
in Figs.6(a) and (b) are defined by 
\eq 
R_{B_{s1} B_s^\ast \gamma} = \frac{1}{16 \pi^2 \, \Lambda^2} \ 
g_{_{B_s^\ast B^\ast K^\ast}} g_{_{B_{s1}}} \  
(g_{_{K^{\ast \pm} K^\pm \gamma}} + g_{_{K^{\ast 0} K^0 \gamma}} ) 
\ I_R \,, 
\en
where $R = G, F$ or $H$ and $I_R$ are the structure integrals given by  
\eq
I_G &=&  
\int\limits_0^\infty\int\limits_0^\infty\int\limits_0^\infty 
\frac{d\alpha_1 d\alpha_2 d\alpha_3}{\Delta_1^3} \, \exp(B_3) \ L_1 \,, 
\nonumber\\
I_F &=&  
\int\limits_0^\infty\int\limits_0^\infty\int\limits_0^\infty 
\frac{d\alpha_1 d\alpha_2 d\alpha_3}{\Delta_1^2} \, \exp(B_3) \ L_2 \,, 
\\ 
I_H &=&  
\int\limits_0^\infty\int\limits_0^\infty\int\limits_0^\infty 
\frac{d\alpha_1 d\alpha_2 d\alpha_3}{\Delta_1^2} \, \exp(B_3) \ L_3 \,, 
\nonumber 
\en 
\eq
B_3 = - \alpha_1 \mu_K^2 - \alpha_2 \mu_{K^\ast}^2 - \alpha_3 \mu_{B^\ast}^2 
- \mu_{B_{s1}}^2 w_{B^\ast K} w_{KB^\ast } 
+ \frac{\alpha_{3KB^\ast}}{\Delta_1} 
(\mu_{B_{s1}}^2 \alpha_{1B^\ast K} + \mu_{B_s^\ast}^2 \alpha_2 ) \,, 
\en 
and 
\eq 
L_1 &=& - \frac{1}{\mu_{B^\ast}^2 (\mu_{B_{s1}}^2 
- \mu_{B_s^\ast}^2)} \biggl( (1 - \beta_{12}) 
(\mu_{B_{s1}}^2 \beta_1 + \mu_{B_s^\ast}^2 (\beta_1 + \beta_2) )  
+ 3 (\mu_{B_{s1}}^2 - \mu_{B_s^\ast}^2) + \frac{3}{\Delta_1} \biggr) 
\,,\nonumber\\ 
L_2 &=& - \biggl(2 + \frac{1}{\mu_{B^\ast}^2 \Delta_1} \biggr) (1 - \beta_1)  
+ \beta_2 \biggl( 3 + \frac{5}{\mu_{B^\ast}^2 \Delta_1} \biggr) 
- \beta_1 \beta_2 \biggl( 1 - \frac{\mu_{B_{s1}}^2 
+ \mu_{B^\ast}^2}{\mu_{B^\ast}^2} + \frac{4}{\mu_{B^\ast}^2 \Delta_1}
\biggr) \nonumber\\
&-&\beta_2^2 \biggl( 1 - 2 \frac{\mu_{B_s^\ast}^2}{\mu_{B^\ast}^2} 
+ \frac{4}{\mu_{B^\ast}^2 \Delta_1}\biggr) 
- \beta_2^3 (3 - 4 \beta_2) \frac{\mu_{B_s^\ast}^2}{\mu_{B^\ast}^2}
-2\beta_1 \beta_2^2 (1 - \beta_2) 
\frac{\mu_{B_{s1}}^2 + 2\mu_{B_s^\ast}^2}{\mu_{B^\ast}^2} \nonumber\\
&-&\beta_1^2\beta_2 (1 - \beta_2) 
\frac{2\mu_{B_{s1}}^2 + \mu_{B_s^\ast}^2}{\mu_{B^\ast}^2} 
+ \beta_1^3 \beta_2 \frac{\mu_{2 B_{s1}}^2}{\mu_{B^\ast}^2} 
\,,\nonumber\\ 
L_3 &=& - \beta_1 (1 - \beta_{12})\,, \hspace*{.25cm} 
\beta_1 \ = \ \frac{\alpha_{1B^\ast K}}{\Delta_1}\,, \hspace*{.25cm} 
\beta_2 \ = \ \frac{\alpha_2}{\Delta_1}\,, \hspace*{.25cm} 
\beta_{12} = \beta_1 + \beta_2 \,. 
\en 

\subsection{Decay $B_{s1}^\ast \to B_{s0}^\ast \gamma$.}

The contribution to the coupling constant 
$G_{B_{s1} B_{s0}^\ast \gamma}$, 
coming from the diagrams 
in Figs.7(a,b) is defined by 
\eq 
G_{B_{s1} B_{s0}^\ast \gamma} = \frac{1}{16 \pi^2 \, \Lambda^2} \,
g_{_{B_{s1}}} g_{_{B_{s0}^\ast}} (g_{_{B^{\ast \pm} B^\pm \gamma}} 
+ g_{_{B^{\ast 0} B^0 \gamma}} ) \ I_{B_{s1} B_{s0}^\ast \gamma} \,, 
\en
where $I_{B_{s1} B_{s0}^\ast \gamma}$ is the structure integral  
given by  
\eq
I_{B_{s1} B_{s0}^\ast \gamma} &=& 
\int\limits_0^\infty\int\limits_0^\infty\int\limits_0^\infty  
\frac{d\alpha_1 d\alpha_2 d\alpha_3}{\Delta^3} \, \exp(B_4) 
(\alpha_3 + w_{B^\ast K} + w_{BK}) \,, \nonumber\\
B_4 &=&  - \alpha_1 \mu_{B^\ast}^2 - \alpha_2 \mu_B^2 
- \alpha_3 \mu_K^2 
- \mu_{B_{s1}}^2 w_{B^\ast K} w_{KB^\ast } 
- \mu_{B_{s0}^\ast }^2 w_{BK} w_{KB} \nonumber\\
&+& \frac{\alpha_3 + w_{B^\ast K} + w_{BK}}{\Delta} 
(\mu_{B_{s1}}^2 \alpha_{1KB^\ast } + \mu_{B_{s0}^\ast}^2 \alpha_{2KB} ) \,, 
\nonumber\\  
\Delta&=& 2 + \alpha_{123} \,. 
\en

\newpage 

\begin{table}
\vspace*{2cm}
\begin{center}
\caption{\label{BsS} Decay widths of $B_{s0}^{\ast} \to B_s \pi$
and $B_{s1} \to B_s^{\ast} \pi^0$ in keV. The range of values for our
results is due to the variation of $\Lambda_M$ from 1 to 2 GeV.}

\vspace*{.25cm}

\def\arraystretch{1.2}
\begin{tabular}{|l|l|l|l|}
\hline \hspace*{.5cm} Approach \hspace*{.5cm} & \hspace*{.5cm}
$\Gamma(B_{s0}^{\ast} \to B_s \pi^0)$ \hspace*{.5cm}
&\hspace*{.5cm} Approach \hspace*{.5cm}& \hspace*{.5cm}
$\Gamma(B_{s1} \to B_s^{\ast} \pi^0)$  \hspace*{.5cm}\\
\hline \,\,\,\,\, Ref.~\cite{Bardeen:2003kt} \,\,\,\,\, &
\,\,\,\,\,\,\,\, 21.5 \,\,\,\,\, &\,\,\,\,\,
Ref.~\cite{Bardeen:2003kt} 
\,\,\,\,\,& \,\,\,\,\,\,\,\, 21.5 \,\,\,\,\, \\ 
\hline \,\,\,\,\, 
Ref.~\cite{Guo:2006fu}
\,\,\,\,\, & \,\,\,\,\,\,\,\, 1.54 \,\,\,\,\, &\,\,\,\,\, 
Ref.~\cite{Guo:2006rp} \,\,\,\,\, & \,\,\,\,\,\,\,\, 10.36 \,\,\,\,\,\\
\hline \,\,\,\,\, 
Ref.~\cite{Wang:2008ny} 
\,\,\,\,\, & \,\,\,\,\,\,\,\, 6.8 $-$ 30.7 \,\,\,\,\, &\,\,\,\,\, 
Ref.~\cite{Wang:2008ny} 
\,\,\,\,\, & \,\,\,\,\,\,\,\, 5.3 $-$ 20.7\,\,\,\,\,\\
\hline \,\,\,\,\, 
Our results & \,\,\,\,\,\,\,\, 55.2 $-$ 89.9 \,\,\,\,\, 
& \,\,\,\,\, 
Our results & \,\,\,\,\,\,\,\, 57.0 $-$ 94.0 \,\,\,\,\,\\
\hline
\end{tabular}
\end{center}
\end{table}

\begin{table}
\vspace*{2cm}
\begin{center}
\caption{\label{BsEM} Decay widths of $B_{s0}^{\ast} \to B_s^{\ast}
\gamma$ and $B_{s1} \to B_s \gamma$ in keV. The range of values for our
results is due to the variation of $\Lambda_M$ from 1 to 2 GeV.}

\vspace*{.25cm}

\def\arraystretch{1.2}
\begin{tabular}{|l|l|l|l|}
\hline \hspace*{.5cm} Approach \hspace*{.5cm} & \hspace*{.5cm}
$\Gamma(B_{s0}^{\ast} \to B_s^{\ast} \gamma)$ \hspace*{.5cm}
&\hspace*{.5cm} Approach \hspace*{.5cm}& \hspace*{.5cm}
$\Gamma(B_{s1} \to B_s \gamma)$ \hspace*{.5cm}\\
\hline \,\,\,\,\, Ref.~\cite{Bardeen:2003kt} \,\,\,\,\, &
\,\,\,\,\,\,\,\, 58.3 \,\,\,\,\, &\,\,\,\,\,
Ref.~\cite{Bardeen:2003kt}
\,\,\,\,\,& \,\,\,\,\,\,\,\, 39.1 \,\,\,\,\, \\
\hline \,\,\,\,\, Ref.~\cite{Vijande:2007ke} \,\,\,\,\, &
\,\,\,\,\,\,\,\, 171.4 \,\,\,\,\, &\,\,\,\,\,
Ref.~\cite{Vijande:2007ke} 
\,\,\,\,\,& \,\,\,\,\,\,\,\, 106.5 \,\,\,\,\, \\
\hline \,\,\,\,\, Ref.~\cite{Vijande:2007ke} \,\,\,\,\, &
\,\,\,\,\,\,\,\, 31.9 \,\,\,\,\, &\,\,\,\,\,
Ref.~\cite{Vijande:2007ke} 
\,\,\,\,\,& \,\,\,\,\,\,\,\, 60.7 \,\,\,\,\, \\
\hline \,\,\,\,\, Ref.~\cite{Wang:2008bt} \,\,\,\,\, &
\,\,\,\,\,\,\,\, 1.3 $-$ 13.6 \,\,\,\,\, &\,\,\,\,\,
Ref.~\cite{Wang:2008bt} 
\,\,\,\,\,& \,\,\,\,\,\,\,\, 3.2 $-$ 15.8 \,\,\,\,\, \\
\hline \,\,\,\,\, 
Our results & \,\,\,\,\,\,\,\, 3.07 $-$ 4.06 \,\,\,\,\, 
& \,\,\,\,\, 
Our results & \,\,\,\,\,\,\,\, 2.01 $-$ 2.67 \,\,\,\,\,\\
\hline
\end{tabular}
\end{center}
\end{table}

\newpage 

\begin{figure}
\begin{center}
\epsfig{file=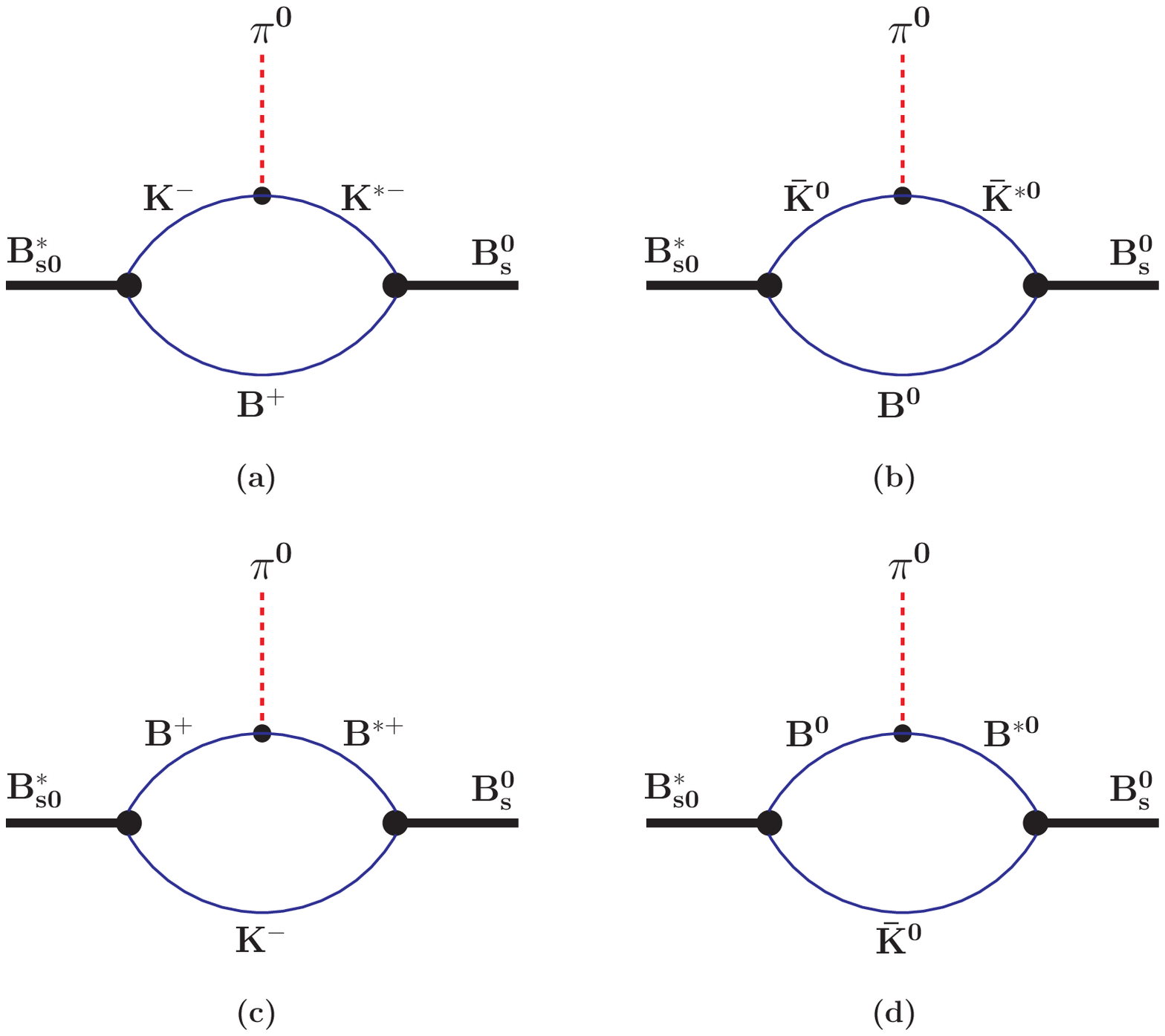, scale=.65}
\caption{Diagrams contributing to the strong transition 
$B_{s0}^\ast \to B_s + \pi^0$.} 
\label{fig:Bs0_str}
\end{center}

\vspace*{.5cm}

\begin{center}
\epsfig{file=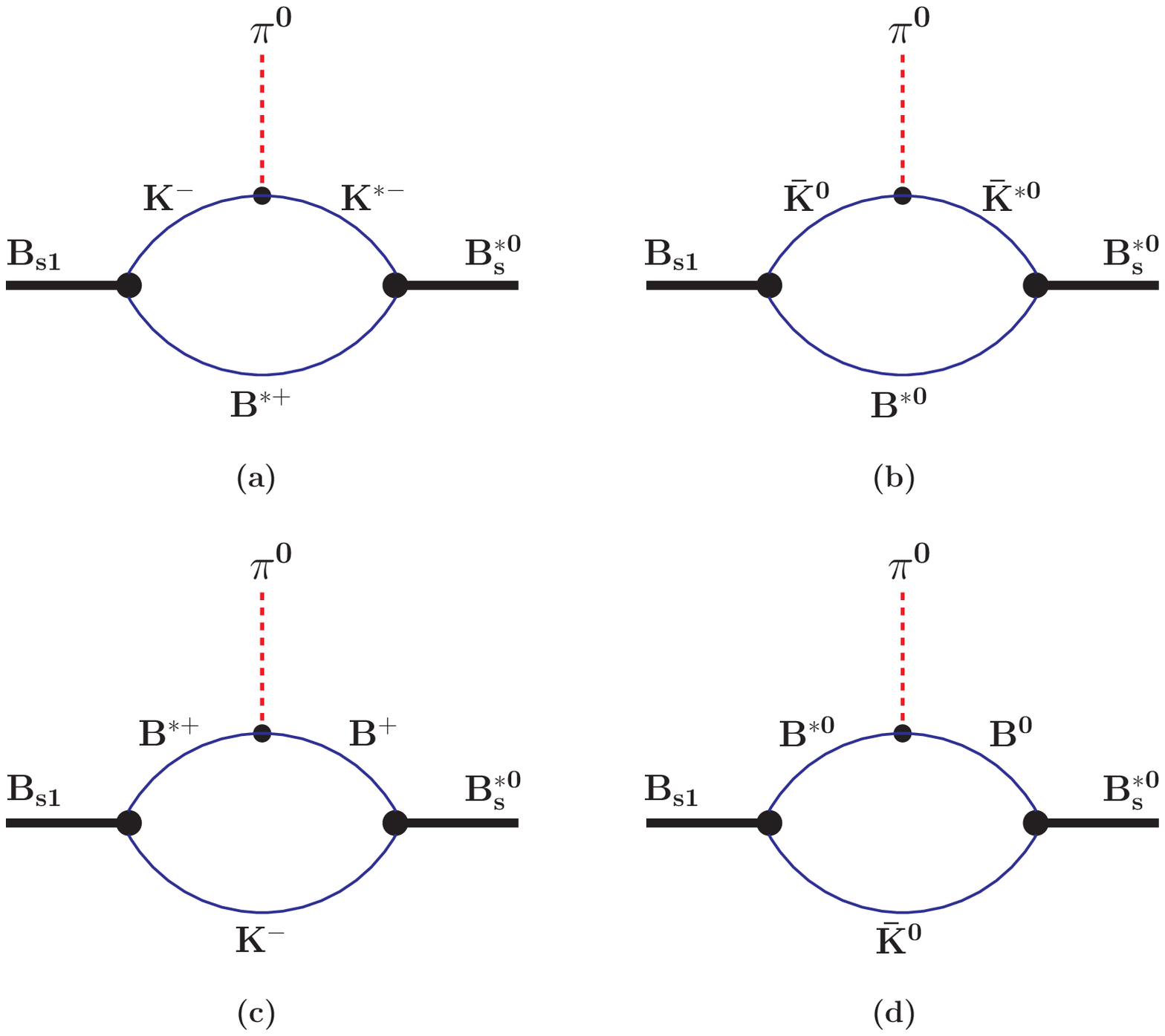, scale=.65}
\caption{Diagrams contributing to the strong transition 
$B_{s1} \to B_s^\ast + \pi^0$.} 
\label{fig:Bs1_str}
\end{center}
\end{figure}

\newpage

\begin{figure}
\begin{center}
\epsfig{file=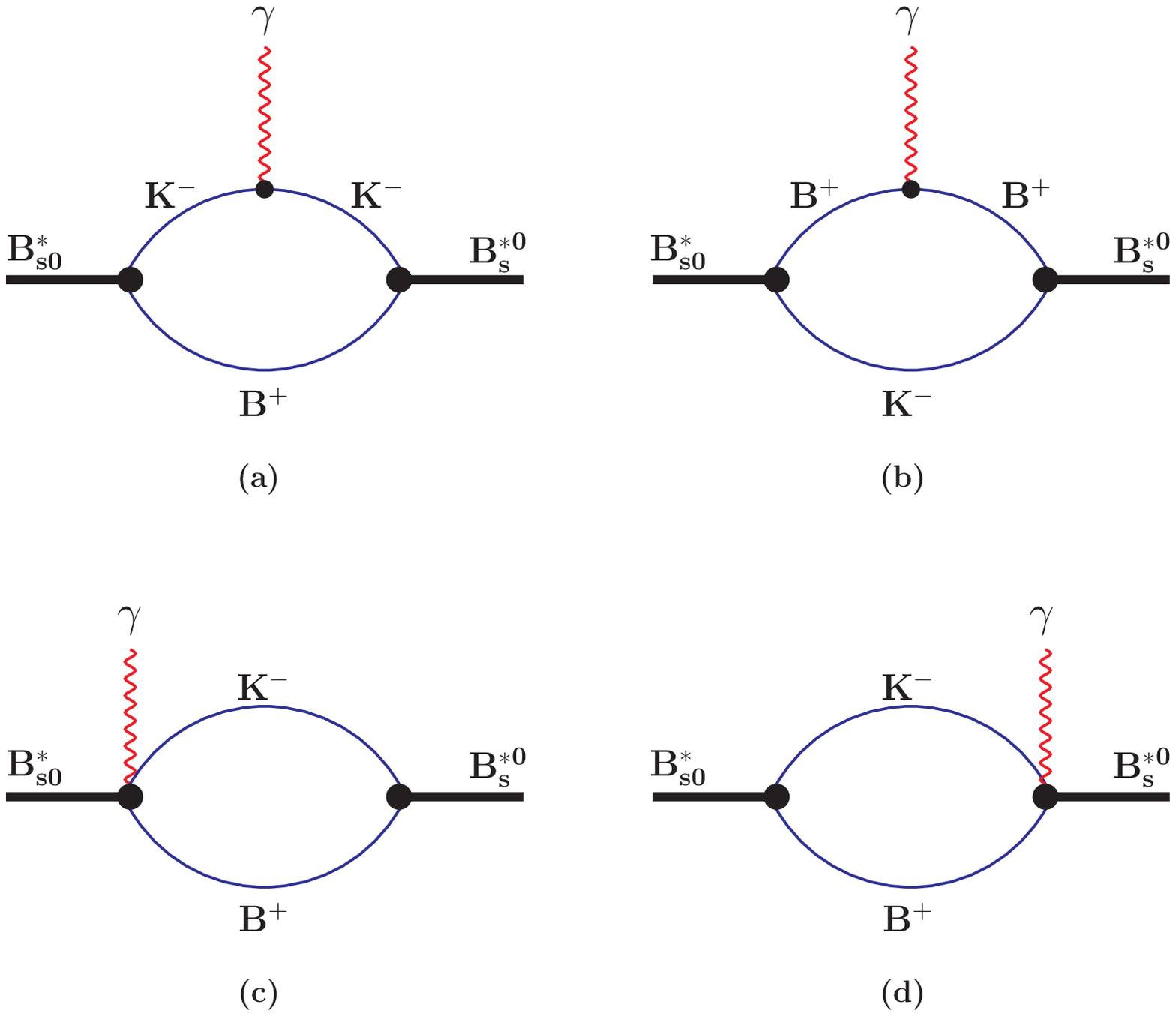, scale=.65}
\end{center}
\caption{Diagrams contributing to the radiative transition
$B_{s0}^\ast \to B_{s}^\ast + \gamma$.}
\label{fig:Bs0_em}

\vspace*{.5cm}

\begin{center}
\epsfig{file=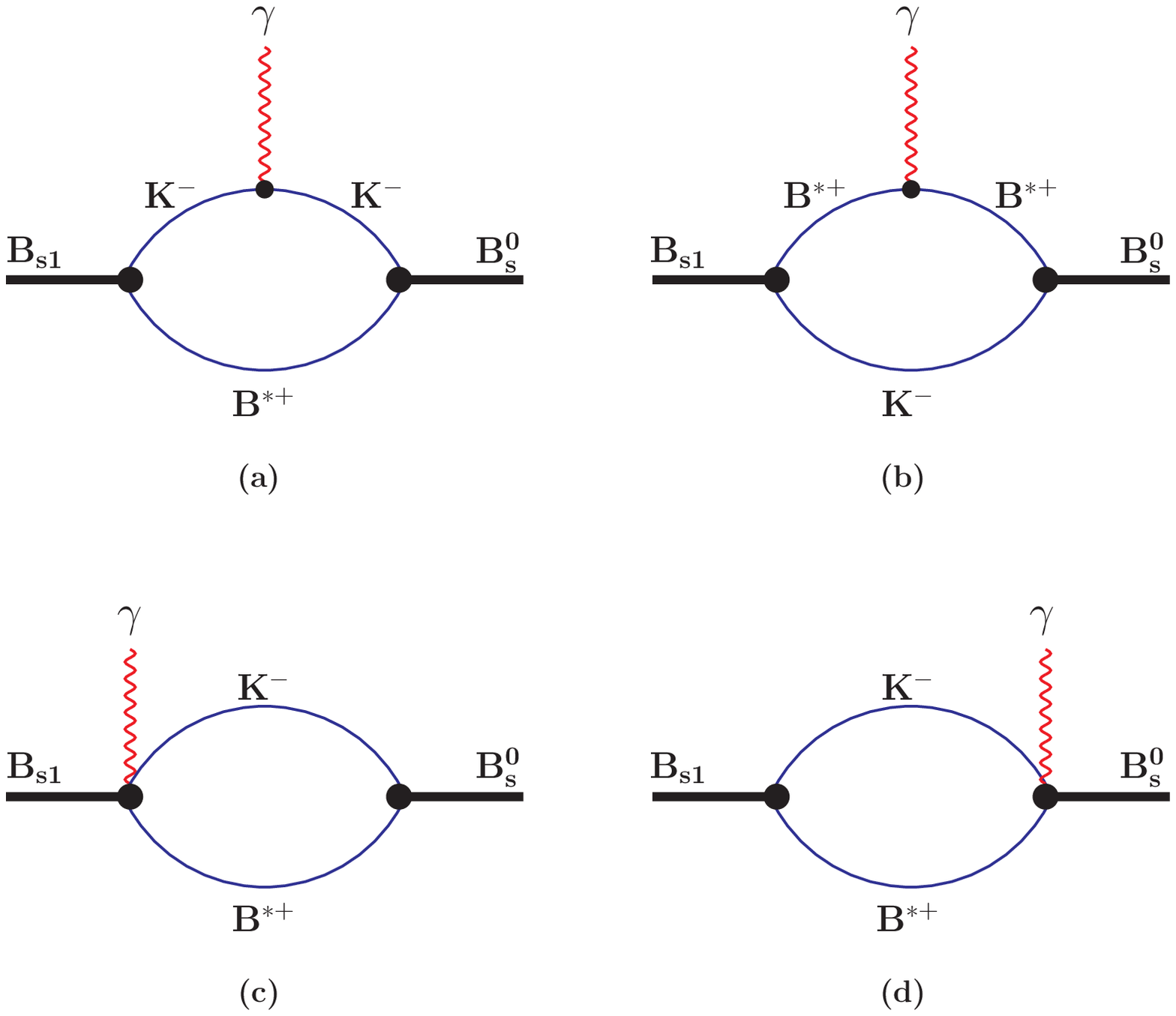, scale=.65}
\caption{Diagrams contributing to the radiative transition
$B_{s1} \to B_s + \gamma$.}
\label{fig:Bs1_em}
\end{center}
\end{figure}

\newpage

\begin{figure}
\begin{center}
\epsfig{file=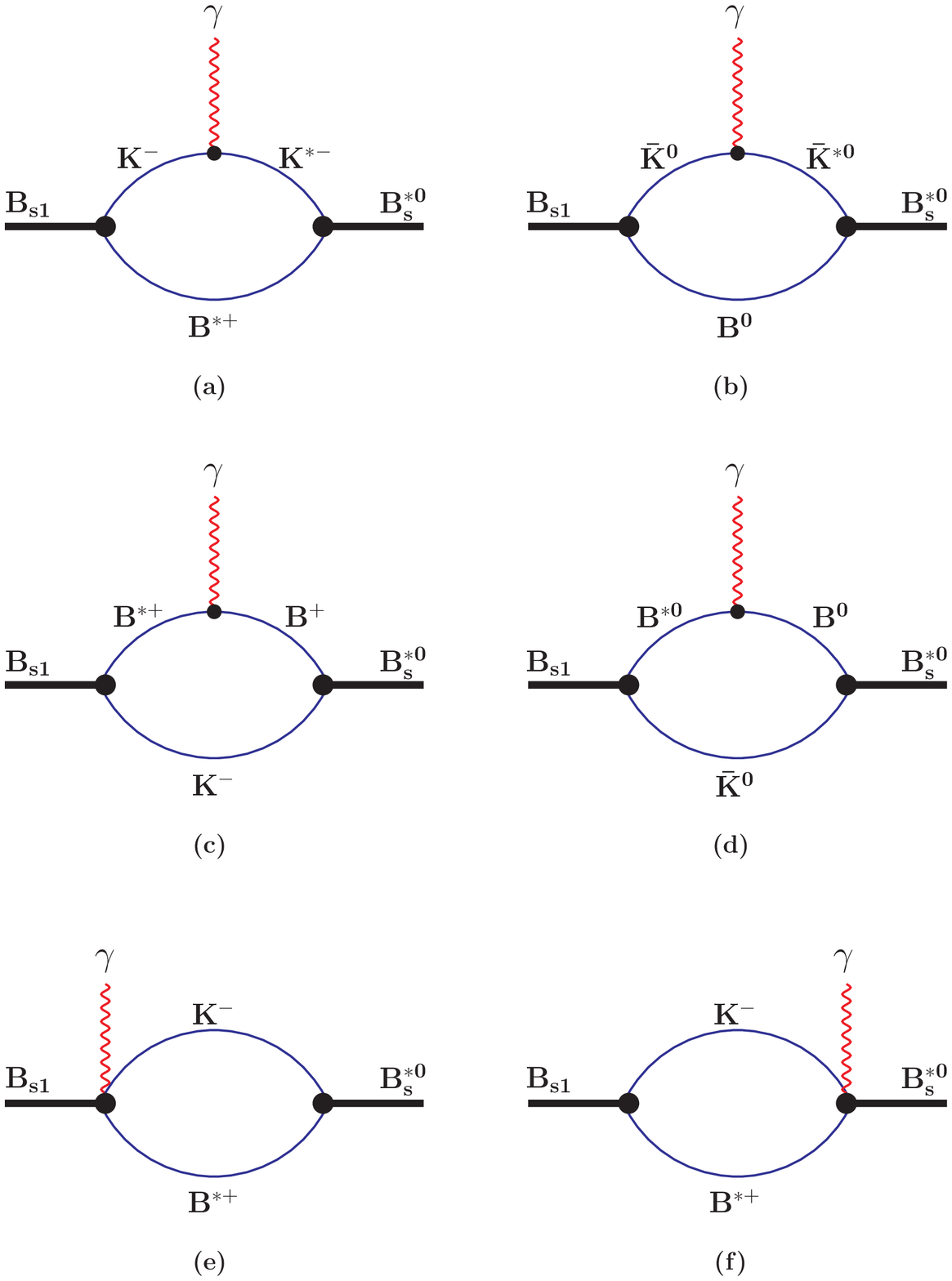, scale=.65}
\end{center}
\caption{Diagrams contributing to the radiative transition
$B_{s1} \to B_{s}^\ast + \gamma$.}
\label{fig:Bs0_em2}

\vspace*{.5cm}

\begin{center}
\epsfig{file=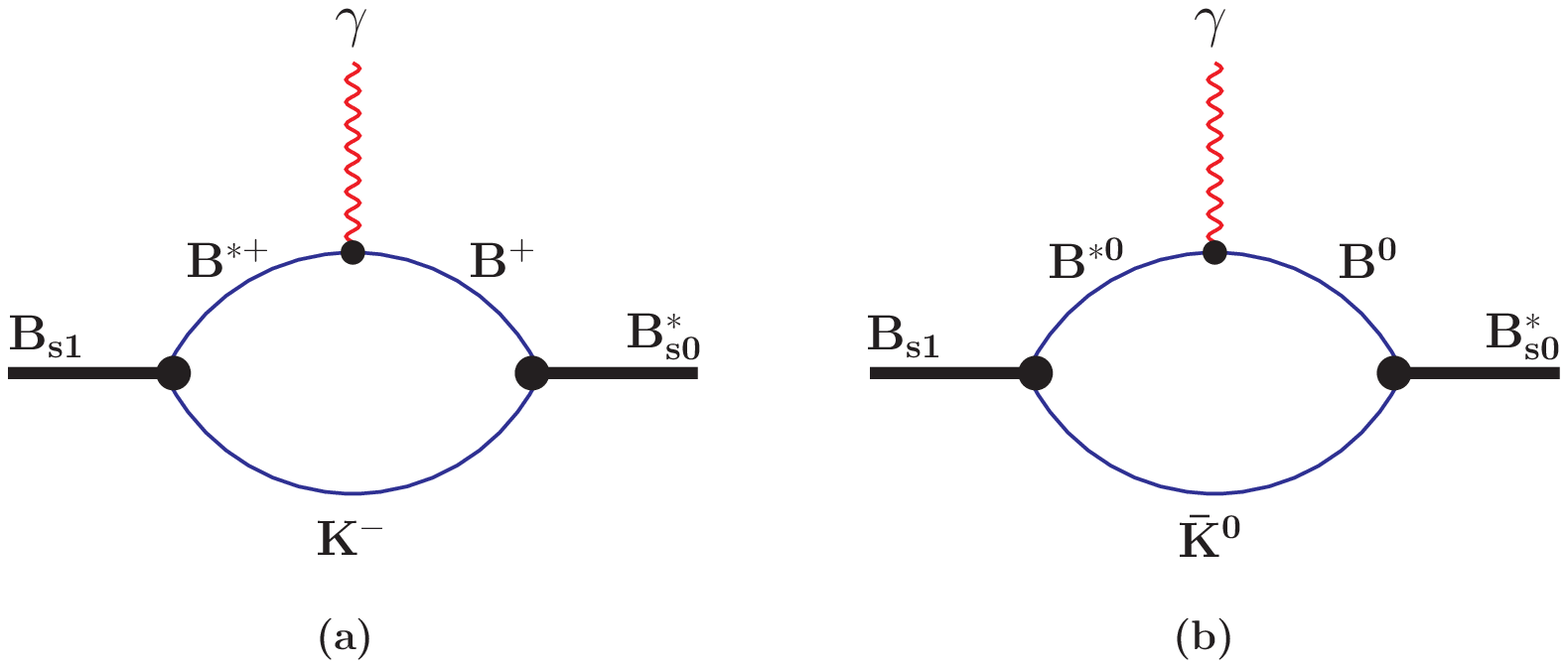, scale=.65}
\caption{Diagrams contributing to the radiative transition
$B_{s1} \to B_{s0}^\ast + \gamma$.}
\label{fig:Bs1_em3}
\end{center}
\end{figure}

\end{document}